\documentclass[12pt,preprint]{aastex}
\usepackage{amssymb}
\def\lambdabar{\lambda\llap {--}}
\shorttitle{Probing MSP light curves using Fermi-LAT}
\shortauthors{Venter et al.}

\begin{document}
\title{PROBING MILLISECOND PULSAR EMISSION GEOMETRY USING LIGHT CURVES FROM THE \textit{FERMI /} LARGE AREA TELESCOPE}

\author{C. VENTER,\altaffilmark{1,2,3} A.~K. HARDING,\altaffilmark{1} AND L. GUILLEMOT\altaffilmark{4}}
\altaffiltext{1}{Astrophysics Science Division, NASA Goddard Space Flight Center, Greenbelt, MD 20771, USA}
\altaffiltext{2}{Unit for Space Physics, North-West University, Potchefstroom Campus, Private Bag X6001, Potchefstroom 2520, South Africa}
\altaffiltext{3}{NASA Postdoctoral Program Fellow}
\altaffiltext{4}{Max-Planck-Institut f\"{u}r Radioastronomie, Auf dem H\"{u}gel 69, 53121 Bonn, Germany}

\begin{abstract}
An interesting new high-energy pulsar sub-population is emerging following early discoveries of gamma-ray millisecond pulsars (MSPs) by the \textit{Fermi} Large Area Telescope (LAT). We present results from 3D emission modeling, including the Special Relativistic effects of aberration and time-of-flight delays and also rotational sweepback of B-field lines, in the geometric context of polar cap (PC), outer gap (OG), and two-pole caustic (TPC) pulsar models. In contrast to the general belief that these very old, rapidly-rotating neutron stars (NSs) should have largely pair-starved magnetospheres due to the absence of significant pair production, we find that most of the light curves are best fit by TPC and OG models, which indicates the presence of narrow accelerating gaps limited by robust pair production -- even in these pulsars with very low spin-down luminosities. The gamma-ray pulse shapes and relative phase lags with respect to the radio pulses point to high-altitude emission being dominant for all geometries. We also find exclusive differentiation of the current gamma-ray MSP population into two MSP sub-classes: light curve shapes and lags across wavebands impose either pair-starved PC (PSPC) or TPC / OG-type geometries. In the first case, the radio pulse has a small lag with respect to the single gamma-ray pulse, while the (first) gamma-ray peak usually trails the radio by a large phase offset in the latter case. Finally, we find that the flux correction factor as a function of magnetic inclination and observer angles is typically of order unity for all models. Our calculation of light curves and flux correction factor for the case of MSPs is therefore complementary to the ``ATLAS paper'' of Watters et al.\ for younger pulsars.
\end{abstract}


\section{Introduction}
\label{sec:intro}
The field of gamma-ray pulsars has already benefited profoundly from discoveries made during the first year of operation of the \textit{Fermi} / Large Area Telescope (LAT). These include detections of the radio-quiet gamma-ray pulsar inside the supernova remnant CTA~1 \citep{Abdo08_CTA1}, the second gamma-ray millisecond pulsar (MSP) \citep{Abdo09_J0030} following the \textit{EGRET} $4.9\sigma$-detection of PSR~J0218+4232 \citep{Kuiper04}, the~6 high-confidence \textit{EGRET} pulsars \citep{Thompson99,Thompson04}, and discovery of~16 radio-quiet pulsars using blind searches \citep{Abdo09_BS}. In addition, 8~MSPs have now been unveiled \citep[][see Table~\ref{tab1}]{Abdo09_MSP}, confirming expectations prior to \textit{Fermi's} launch in June 2008 \citep{HUM05,Venter05_Cherenkov}. A \textit{Fermi} six-month pulsar catalog is expected to be released shortly \citep{Abdo09_Cat}. \textit{AGILE} has also reported the discovery of~4 new gamma-ray pulsars, and marginal detection of~4 more \citep{Halpern08,Pellizzoni09b}, in addition to the detection of~4 of the \textit{EGRET} pulsars \citep{Pellizzoni09a}. Except for the detection of the Crab at energies above 25~GeV \citep{Aliu08}, no other pulsed emission from pulsars has as yet been detected by ground-based Cherenkov telescopes \citep{Schmidt05,Albert07,Aharonian07,Fuessling08,Kildea08,Konopelko08,Albert08,Celik08,DelosReyes09}.

\begin{figure}
\epsscale{0.8}
\plotone{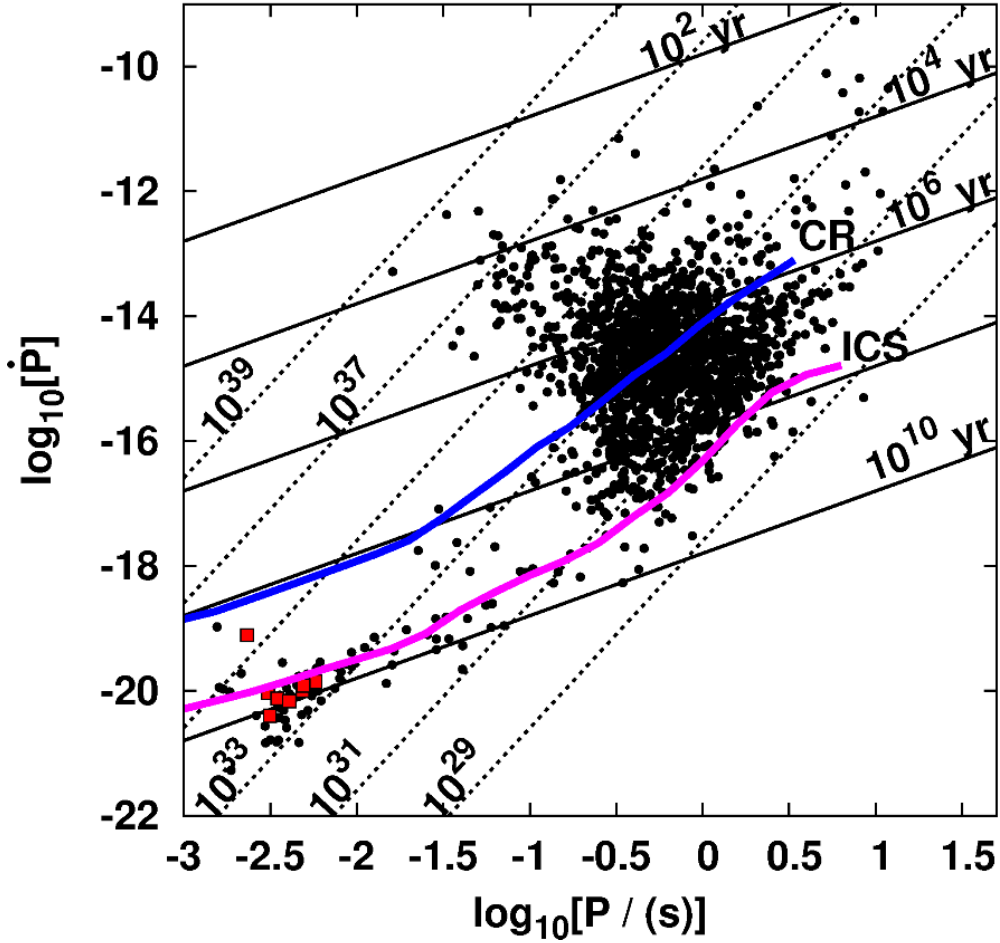}
\caption{The $P\dot{P}$-diagram, indicating contours of constant $\dot{E}_{\rm rot}$ (dashed lines) and rotational age (solid lines), as well as pulsars from the ATNF Catalog \citep{ATNF}. We used values of $\dot{P}>0$ corrected for the Shklovskii effect \citep{Shklovskii70}, and removed pulsars in globular clusters. The squares are the 8 newly-discovered \textit{Fermi} MSPs \citep{Abdo09_MSP}. All except PSR~J0218+4232 lie below the ICS deathline, and all eight lie below the CR deathline \citep[modeled by][]{HM02}.\label{fig:PPdot}}
\end{figure}

MSPs are characterized by relatively short periods $P\lesssim30$~ms and low surface magnetic fields $B_0\sim10^8-10^9$~G, and appear in the lower left corner of the $P\dot{P}$-diagram (with $\dot{P}$ the time-derivative of $P$; see Figure~\ref{fig:PPdot}, where the newly-discovered \textit{Fermi} MSPs are indicated by squares). MSPs are thought to have been spun-up to millisecond periods by transfer of mass and angular momentum from a binary companion during an accretion phase \citep{Alpar82}. This follows an evolutionary phase of cessation of radio emission from their mature pulsar progenitors, after these have spun down to long periods and crossed the ``death line'' for radio emission. These ``radio-silent'' progenitors \citep{Glendenning00} are thought to reside in the ``death valley'' of the $P\dot{P}$-diagram, which lies below the inverse Compton scattering (ICS) pair death line \citep{HM02}.

The standard ``recycling scenario'' \citep{Bhattacharya91} hypothesizing that MSP birth is connected to low-mass X-ray binaries (LMXRBs) might have been confirmed recently by the detection of radio pulsations from a nearby MSP in an LMXRB system, with an optical companion star \citep{Archibald09}. Optical observations indicate the presence of an accretion disk within the past decade, but none today, raising the possibility that the radio MSP has ``turned on'' after termination of recent accretion activity, thus providing a link between LMXRBs and the birth of radio MSPs.

High-energy (HE) radiation from pulsars has mainly been explained as originating from two emission regions. Polar cap (PC) models \citep{Harding78,Daugherty82,Sturner95,DH96} assume extraction of primaries from the stellar surface and magnetic pair production of ensuing HE curvature radiation (CR) or ICS gamma rays, leading to low-altitude pair formation fronts (PFFs) which screen the accelerating electric field \citep{HM98,HM01,HM02}. These space-charge-limited-flow (SCLF) models have since been extended to allow for the variation of the CR PFF altitude across the PC and therefore acceleration of primaries along the last open magnetic field lines in a slot gap (SG) scenario \citep{AS79,Arons83,MH03_SG,MH04_SG,Harding08_Crab}. The SG results from the absence of pair creation along these field lines, forming a narrow acceleration gap that extends from the neutron star (NS) surface to near the light cylinder. The SG model is thus a possible physical realization of the two-pole caustic (TPC) geometry \citep{Dyks03}, developed to explain pulsar HE light curves. On the other hand, outer gap (OG) models \citep{CHR86a,CHR86b,CR92,CR94,Romani96,Cheng00,Zhang04} assume that HE radiation is produced by photon-photon pair production-induced cascades along the last open field lines above the null-charge surfaces ($\mathbf{\Omega}\cdot\mathbf{B}=0$, with $\Omega=2\pi/P$), where the Goldreich-Julian charge density \citep{GJ69} changes sign. The pairs screen the accelerating E-field, and limit both the parallel and transverse gap size \citep{Takata04}. Classical OG models may be categorized as ``one-pole caustic models'', as the assumed geometry prevents observation of radiation from gaps (caustics) associated with both magnetic poles \citep{Harding05}. More recently, however, \citet{Hirotani06,Hirotani07} found and applied a 2D, and subsequently a 3D \citep{Hirotani08} OG solution which extends toward the NS surface, where a small acceleration field extracts ions from the stellar surface in an SCLF-regime (see also \citet{Takata04,Takata06}, and in particular \citet{Takata08} for application to Vela). Lastly, \citet{Takata09} modeled Geminga using an OG residing between a ``critical'' B-field line (perpendicular to the rotational axis at the light cylinder) and the last open field line.

Current models using dipole field structure to model MSPs predict largely unscreened magnetospheres due to the relatively low B-fields inhibiting copious magnetic pair production. Such pulsars may be described by a variation of the PC model (applicable for younger pulsars), which we will refer to as a ``pair-starved polar cap'' (PSPC) model \citep{MH04_PS,HUM05,MH09_PS}. In a PSPC model, the pair multiplicity is not high enough to screen the accelerating electric field, and charges are continually accelerated up to high altitudes over the full open-field-line region. The formation of a PSPC ``gap'' is furthermore naturally understood in the context of an SG accelerator progressively increasing in size with pulsar age, which, in the limit of no electric field screening, relaxes to a PSPC structure.

Several authors have modeled MSP gamma-ray fluxes, spectra and light curves in both the PSPC \citep{Frackowiak05a,Frackowiak05b,HUM05,VdeJ05,Venter_phd,Zajczyk08} and OG \citep{Zhang03,Zhang07} cases. Collective emission from a population of MSPs in globular clusters \citep{HUM05,Zhang07,BS07,Venter08,Venter09} and in the Galactic Center \citep{Wang06} have also been considered. \citet{Watters09} recently calculated beaming patterns and light curves from a population of canonical pulsars with spin-down luminosities $\dot{E}_{\rm rot} > 10^{34}$~erg\,s$^{-1}$ using geometric PC, TPC, and OG models. They obtained predictions of peak multiplicity, peak separation, and flux correction factor $f_\Omega$ as functions of magnetic inclination and observer angles $\alpha$ and $\zeta$, and gap width $w$. The latter factor $f_\Omega$ is used for converting observed phase-averaged energy flux $G_{\rm obs}$ to the total radiated (gamma-ray) luminosity $L_\gamma$, which is important for calculating the efficiency of converting $\dot{E}_{\rm rot}$ into $L_\gamma$. A good example is the inference of the conversion efficiencies of globular-cluster MSPs which may be collectively responsible for the HE radiation observed from 47~Tucanae by \textit{Fermi}-LAT \citep{Abdo09_Tuc}.

In this paper, we present results from 3D emission modeling, including Special Relativistic (SR) effects of aberration and time-of-flight delays, and rotational sweepback of B-field lines, in the geometric context of OG, TPC, and PSPC pulsar models. We study the newly-discovered gamma-ray MSP population \citep{Abdo09_MSP}, and obtain fits for gamma-ray and radio light curves. Our calculation of light curves and flux correction factors $f_\Omega(\alpha,\zeta,P)$ for the case of MSPs is therefore complementary to the work of \citet{Watters09} which focuses on younger pulsars, although our TPC and OG models include non-zero emission width. Section~\ref{sec:Model} deals with details of the various models we have applied. We discuss light curves from both observational and theoretical perspectives in Section~\ref{sec:LCs}, and present our results and conclusions in Sections~\ref{sec:Results} and~\ref{sec:Con}. 

\section{Model Description}
\label{sec:Model}

\subsection{B-field and SR Effects}
\label{sec:B}
\citet{Deutsch55} found the solution of the B- and E-fields exterior to a perfectly-conducting sphere which rotates in vacuum as an inclined rotator. We assume that this retarded vacuum dipolar B-field is representative of the magnetospheric structure, and we use the implementation by \citet{Dyks04_B,Dyks04_Bovc}, following earlier work by \citet{Romani95,Higgins97,Arendt98,Cheng00}. For this B-field, the PC shape is distorted asymmetrically by rotational sweepback of field lines. Each field line's footpoint is labeled by the open volume coordinates $(r_{\rm ovc},l_{\rm ovc})$ as defined by \citet{Dyks04_B}, with $r_{\rm ovc}$ labeling self-similar contours or ``rings'' ($r_{\rm ovc}$ is normalized to the PC radius $R_{\rm PC}$), and $l_{\rm ovc}$ giving the arclength along a ring (analogous to azimuthal angle; also refer to \citet{Harding08_Crab} for more details). 

We calculate the rim of the PC by tracing field lines which close at the light cylinder back to the stellar surface, and then divide this PC into rings \citep[see e.g.\ Figure~2 of][]{Dyks04_Bovc} and azimuthal bins, with each surface patch $dS$ associated with a particular B-field line. We follow primary electrons moving along each field line, and collect radiation (corrected for SR-effects) in a phaseplot map (Section~\ref{sec:Phaseplots}). Following \citet{CR92,Cheng00,Dyks03}, we assume constant emissivity along the B-lines in the gap regions of the geometric PC, OG, and TPC models (but not for the PSPC model), so that we do not need to include any particular E-field (or calculate $dS$ explicitly) for these. In the case of the PSPC model, we use the approximation $\xi\approx r_{\rm ovc}$ (with $\xi \equiv \theta/\theta_{\rm pc}$ the normalized polar angle, and $\theta_{\rm pc}\approx(\Omega R/c)^{1/2}$ the PC angle), and include the full E-field up to high altitudes (Section~\ref{sec:Transport}).

In addition to the rotational sweepback (retardation) of the B-lines, we include the effects of aberration and time-of-flight delays. We calculate the position and direction of photon propagation (assumed to be initially tangent to the local B-line) in the co-rotating frame, and then aberrate this direction using a Lorentz transformation, transforming from the instantaneously co-moving frame to the IOF. Lastly, we correct the phase at which the photon reaches the observer for time delays due to the finite speed of light. More details about calculation of these SR effects may be found in \citet{Dyks04_SR,Dyks04_Bovc}, following previous work by e.g., \citet{Morini83,Romani95}. We furthermore explicitly use the curvature radius of the B-field lines as calculated in the inertial observer frame (IOF), and not in the co-rotating frame, when performing particle transport calculations (Section~\ref{sec:Transport}). Such a model has also recently been applied to the Crab by \citet{Harding08_Crab}.

We have calculated TPC and OG models assuming gaps that are confined between two B-field lines with footpoints at $r_{\rm ovc, 1}$ and $r_{\rm ovc, 2}$. We therefore activated only a small number of rings near the rim ($r_{\rm ovc}\sim1$) with $r_{\rm ovc} \in [r_{\rm ovc, 1},r_{\rm ovc, 2}]$, and binning radiation from these, assuming constant emissivity over the emitting volume.  For TPC models, we used $r_{\rm ovc} \in [0.80,1.00]$, $[0.60,1.00]$, $[0.90,1.00]$, $[0.95,1.00]$, and $[1.00,1.00]$ (see Table~\ref{tab3}) corresponding to gap widths of $w\equiv r_{\rm ovc, 2}-r_{\rm ovc, 1}=0.20$, 0.40, 0.10, 0.05, and 0.00 (the last of these is what we have referred to as the TPC model). Similarly, we investigated OG models with $r_{\rm ovc} \in [0.90,0.90]$, $[1.00,1.00]$, $[0.95,1.00]$ (widths of $w=0.00$, 0.00, and 0.05). These widths are smaller than e.g.\ the value of $\sim0.14$ used by \citet{Hirotani08_apjl}. We did not find good light curve fits for TPC models with large $w$. In the case of OG models, one should consider non-uniform emission when choosing large $w$, which is beyond the scope of this paper. The assumption of constant emissivity in the emitting volume is a simplification, as OG models are expected to produce the bulk of the gamma-radiation along the inner edge ($r_{\rm ovc,inner} < r_{\rm ovc} < r_{\rm ovc,PFF}$) of the gap ($r_{\rm ovc,PFF}<r_{\rm ovc}<1$), with $r_{\rm ovc,PFF}$ indicating the position of the PFF, and $r_{\rm ovc,inner}$ some smaller radius depending on the radiation surface thickness \citep{Watters09}. We lastly modeled the PC and PSPC cases with $r_{\rm ovc} \in [0.00,1.00]$ (i.e., the full open-field-line volume, for both constant emissivity and full radiation codes). We used 180 colatitude ($\zeta$) and phase ($\phi$) bins and individual ring separations of $\delta r_{\rm ovc} = 0.005$, while collecting all photons with energies above 100~MeV (in the case of the PSPC model) when producing phaseplots and subsequent light curves.

It is important to note a critical difference between the radiation distribution in our TPC and OG models and that of \citet{Watters09}. We assume that emission is distributed uniformly throughout the gaps between $r_{\rm ovc, 1}$ and $r_{\rm ovc, 2}$, so the radiation originates from a volume with non-zero width across field lines and the radiation and gap widths are the same. For the TPC model, this geometry is similar to that adopted by \citet{Dyks04_B}, although \citet{Dyks04_B} assummed a Gaussian distribution of emission centered at the gap midpoint while we simply assume a constant emissivity across the gap, both of which crudely approximate the radiation pattern expected in the SG. \citet{Watters09} assume that the emission occurs only along the inner edge of both the TPC and OG gaps ($r_{\rm ovc, 1}$ in our notation), and so their radiation width is confined to a single field line and not equal to their gap width ($w$ in their notation). In the case of the OG, the physically realistic emission pattern would have a non-zero width lying somewhere between infinitely thin and uniform assumptions \citep[see][]{Hirotani08}.

\subsection{Particle Transport and PSPC E-field}
\label{sec:Transport}
We only consider CR losses suffered by electron primaries moving along the B-field lines when modeling the HE emission. In this case, the (single electron) transport equation is given by \citep[e.g.,][]{Sturner95a,DH96}
\begin{equation}
 \dot{E}_{\rm e} = \dot{E}_{\rm e,gain} + \dot{\gamma}_{_{\rm CR}}m_{\rm e}c^2 = e\beta_r cE_{||} - \frac{2e^2c}{3\rho_c^2}\beta_r^4\gamma^4,\label{eq:transport}
\end{equation}
where $c$ is the speed of light in vacuum, $\beta_r=v_r/c\sim1$ the particle velocity, $e$ is the electron charge, $\gamma$ is the electron Lorentz factor, $\dot{\gamma}_{_{\rm CR}}m_ec^2$ the frequency-integrated (total) CR loss rate per particle \citep{BRD00}, $\rho_c$ the curvature radius (as calculated in the IOF; see Section~\ref{sec:B}), and $E_{||}$ the accelerating E-field parallel to the B-field. The acceleration and loss terms balance at a particular $\gamma_{\rm RR}$ in the radiation reaction regime \citep{Luo00}:
\begin{equation}
 \gamma_{\rm RR} = \left(\frac{3 E_{||}\rho_c^2}{2e\beta_r^3}\right)^{1/4}.
\end{equation}

Previous studies \citep[e.g.,][]{VdeJ05,Frackowiak05a,Frackowiak05b,HUM05,Venter08,Zajczyk08} have used the solutions of \citet{MH97,HM98} for the PSPC E-field:
\begin{eqnarray}
E_{||}^{(1)} & = & -\frac{\Phi_0}{R}\left(\theta^{\rm GR}_0\right)^2\left\{12\kappa^\prime s_1\cos\alpha + 6s_2\theta^{\rm GR}_0H(1)\delta^\prime(1)\sin\alpha\cos\phi_{\rm pc}\right\}\\
E_{||}^{(2)} & = & -\frac{\Phi_0}{R}\left(\theta^{\rm GR}_0\right)^2\left\{\frac{3}{2}\frac{\kappa^\prime}{\eta^4}\cos\alpha + \frac{3}{8}\theta^{\rm GR}(\eta)H(\eta)\delta^\prime(\eta)\xi\sin\alpha\cos\phi_{\rm pc}\right\}\left(1-\xi^2\right),\label{eq:E2}\\
\end{eqnarray}
with
\begin{eqnarray}
\Phi_0 & \equiv & \frac{B_0\Omega R^2}{c},\\
\epsilon & \equiv & \frac{2GM}{c^2R},\\
\theta^{\rm GR}(\eta) & \approx & \left(\frac{\Omega R}{c}\frac{\eta}{f(\eta)}\right)^{1/2} \approx \theta_{\rm pc}\\
s_1 & = & \sum_{i=1}^{\infty}\frac{J_0(k_i\xi)}{k_i^3J_1(k_i)}\mathcal{F}_1(\gamma_i(1),\eta)\\
s_2 & = & \sum_{i=1}^{\infty}\frac{J_1(\tilde{k}_i\xi)}{\tilde{k}_i^3J_2(\tilde{k}_i)}\mathcal{F}_1(\tilde{\gamma}_i(1),\eta)\\
\gamma_i(\eta) & = & \frac{k_i}{\eta\theta^{\rm GR}(\eta)(1-\epsilon/\eta)^{1/2}}\label{eq:gammai}\\
\tilde{\gamma}_i(\eta) & = & \frac{\tilde{k}_i}{\eta\theta^{\rm GR}(\eta)(1-\epsilon/\eta)^{1/2}}\label{eq:gammai_tilde}\\
\mathcal{F}_1(\gamma,\eta) & = & 1-e^{-\gamma(1)(\eta-1)},
\end{eqnarray}
and $k_i$ and $\tilde{k}_i$ are the positive roots of the Bessel functions $J_0$ and $J_1$ (with $k_{i+1}>k_i$ and $\tilde{k}_{i+1}>\tilde{k}_i$); $\theta^{\rm GR}_0\equiv\theta^{\rm GR}(1)$; $\gamma(1)$ may be $\gamma_i(1)$ or $\tilde{\gamma}_i(1)$ in the expression for $\mathcal{F}_1$. The functions $H(\eta)$, $f(\eta)$, and $\delta^\prime(\eta)$ are all of order unity, and are defined in \citet{MT92}. The first solution $E_{||}^{(1)}$ is valid for $\eta-1\ll1$, and $E_{||}^{(2)}$ for $\theta^{\rm GR}_0 \ll \eta - 1 \ll c/(\Omega R)$; $R$ is the stellar radius, $\eta=r/R$, $\alpha$ the angle between the rotation and magnetic axes, $\phi_{\rm pc}$ the magnetic azimuthal angle, $\kappa^\prime=2GI/(c^2R^3)$ the General Relativistic (GR) inertial frame-dragging factor (distinct from the $\kappa(x)$ function to be defined later), and $I$ the moment of inertia. 

\citet{MH04_PS} found the solution of $E_{||}$ for altitudes close to the light cylinder in the small-angle approximation (small $\alpha$, $\xi$, and high altitude):
\begin{eqnarray}
E_{||}^{(3)} & \approx & -\frac{3}{16}\left(\frac{\Omega R}{c}\right)^3\frac{B_0}{f(1)}\left[\kappa^\prime\left(1-\frac{1}{\eta_{\rm c}^3}\right)\left(1+\xi^2\right)\cos\alpha\right.\nonumber\\
& +&\frac{1}{2}\left(\sqrt{\eta_{\rm c}}-1\right)\left(\frac{\Omega R}{c}\right)^{1/2}\lambda\left(1+2\xi^2\right)\nonumber\\
& \times&\left.\xi\sin\alpha\cos\phi_{\rm pc}\right]\left(1-\xi^2\right),
\end{eqnarray}
and $\lambda$ is defined after Eq.~(35) of \citet{MH04_PS}. They proposed that one should employ the following formula to match the last two solutions:
\begin{equation}
 E_{||} \approx E_{||}^{(2)}\exp\left[-(\eta-1)/(\eta_{\rm c}-1)\right]+E_{||}^{(3)},
\end{equation}
with $\eta_{\rm c}$ a radial parameter to be determined using a matching procedure. \citet{MH04_PS} estimated that $\eta_{\rm c}\sim3-4$ for MSPs when $\xi=\theta/\theta^{\rm GR}_0\sim0.5$. 

It is important to include the high-altitude solution $E_{||}^{(3)}$, as \textit{Fermi} results seem to indicate that the HE radiation is originating in the outer magnetosphere \citep[e.g.,][]{Abdo09_MSP}. Beaming properties and spectral characteristics of the emission may therefore be quite different in comparison to calculations which only employ $E_{||}^{(1)}$ and $E_{||}^{(2)}$. In addition, while we use E-field expressions derived in the small-angle approximation, it is preferable to use the full solution of the Poisson equation, particularly in the case of MSPs which have relatively small magnetospheres and therefore much larger PC angles compared to canonical pulsars.

\begin{figure}
\epsscale{0.8}
\plotone{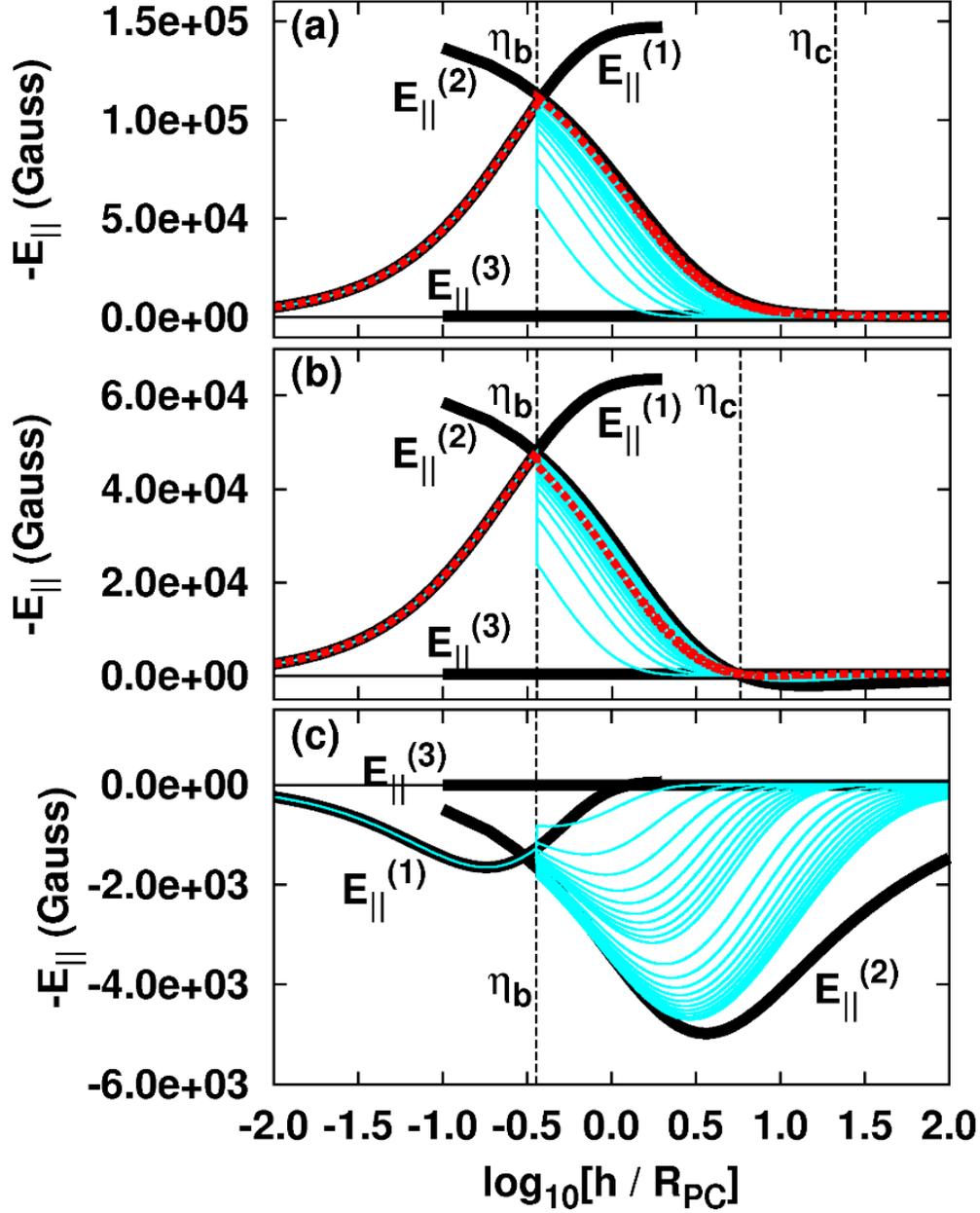}
\caption{Examples of the final E-field we obtain after matching $E_{||}^{(1)}$ through $E_{||}^{(3)}$ for different parameters: $-E_{||}$ vs.\ $\log_{10}$ of the height above the PC, normalized by the PC radius $R_{\rm PC} = (\Omega R^3/c)^{1/2}$. These plots were obtained for $P=5.75\times10^{-3}$~s, $\dot{P}=10^{-20}$, $R=10^6$~cm, and $M=1.4M_\odot$. In panel~(a), we chose~$\alpha=20^\circ$, $\xi=0.3$, $\phi_{\rm pc}=45^\circ$, in panel~(b), $\alpha=35^\circ$, $\xi=0.7$, $\phi_{\rm pc}=150^\circ$, and in panel~(c), $\alpha=80^\circ$, $\xi=0.8$, $\phi_{\rm pc}=200^\circ$. In the last panel, the final $-E_{||}$ is negative, so no solution of $\eta_{\rm c}$ is obtained. In each case, we label $E_{||}^{(1)}$ through $E_{||}^{(3)}$ (thick solid lines), indicate potential solutions (which vary with $\eta_{\rm c}$) by thin gray (cyan) lines, and the final solution by thick (red) dashed lines. Also, we indicate $\eta_{\rm b}$ where we match $E_{||}^{(1)}$ and $E_{||}^{(2)}$, and $\eta_{\rm c}$ where we match $E_{||}^{(2)}$ and $E_{||}^{(3)}$, by thin vertical dashed lines. (Although $E_{||}^{(3)}$ does slightly vary with $\eta_{\rm c}$, we only indicate the $E_{||}^{(3)}$-solution corresponding to the $\eta_{\rm c}$ found for the final solution. For panel~(c), we show a typical $E_{||}^{(3)}$-solution.)\label{fig:etac}}
\end{figure}

In this paper, we calculate $\eta_{\rm c}(P,\dot{P},\alpha,\xi,\phi_{\rm pc})$ explicitly for each B-field line according to the following criteria (we use $\dot{P}=10^{-20}$, $M=1.4M_\odot$, $R=10^6$~cm, and $I=0.4MR^2$ throughout). We require that the resulting E-field should:\\
(1) Be negative for all $1 \leq \eta\lesssim c/(\Omega R)$;\\
(2) Match the part of the $E_{||}^{(2)}$-solution which exceeds $E_{||}^{(3)}$ in absolute magnitude (i.e., where $-E_{||}^{(2)}>-E_{||}^{(3)}$) as closely as possible;\\
(3) Tend toward $E_{||}^{(3)}$ for large $\eta$.\\
The first criterion is required to mitigate the problem of particle oscillations which occurs when the E-field reverses sign beyond some altitude. Instead of this happening $\sim40$\% of the time \citep{VdeJ05,Venter_phd}, we now only have to ignore solutions where $-E_{||}<0$ for $\eta>1.1$ for $\sim8$\% of the time. 
The two lower-altitude solutions $E_{||}^{(1)}$ and $E_{||}^{(2)}$ have been matched at $\eta=\eta_{\rm b}$, using \citep{Venter_phd}
\begin{equation}
 \eta_{\rm b} \approx 1 + 0.0123P^{-0.333}.
\end{equation}
Example fits for $E_{||}$ are shown in Figure~\ref{fig:etac} for different parameters, as noted in the caption. The top two panels show fits for two different $\eta_{\rm c}$, while the bottom panel is an example where no solution for $\eta_{\rm c}$ is found (according to the first criterion above).

\begin{figure}
\epsscale{0.8}
\plotone{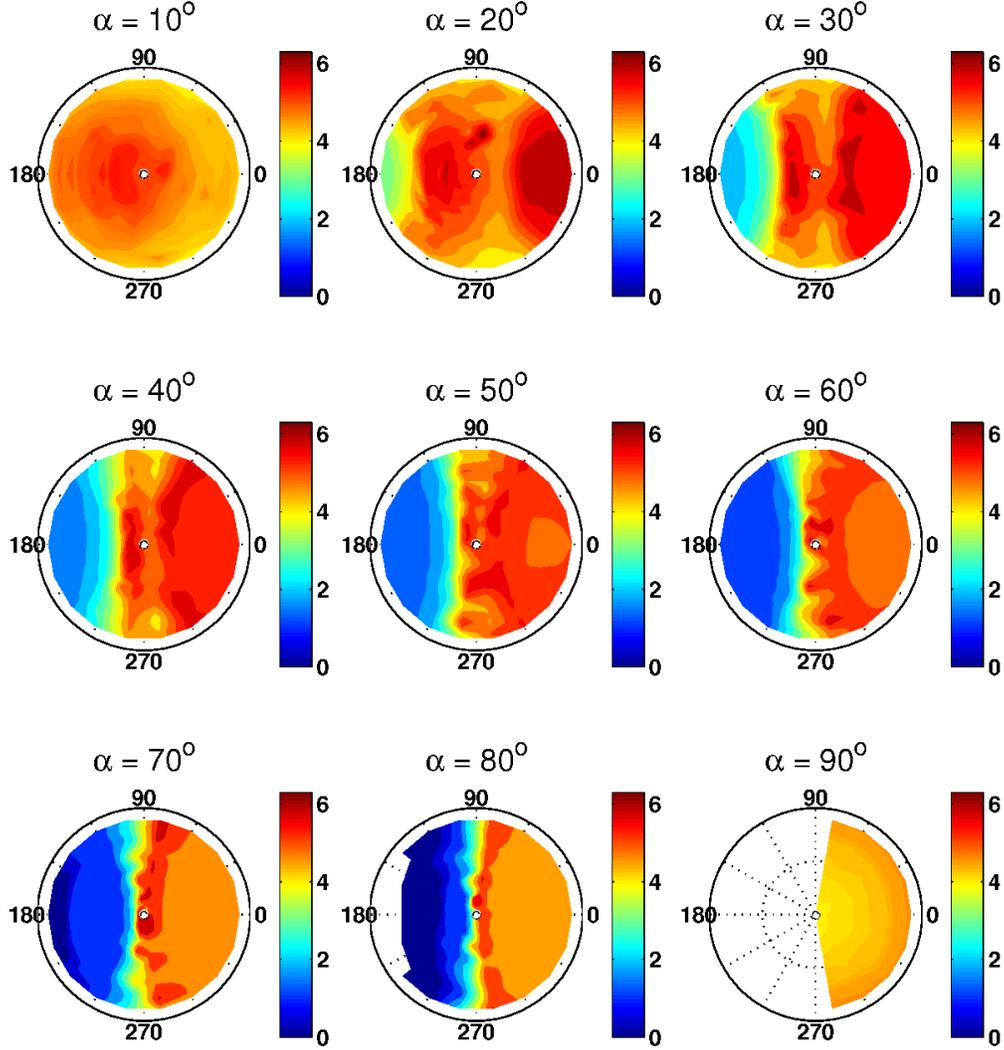}
\caption{Contour plots of our solutions of $\eta_{\rm c}$ for $P=5$~ms, and for $\alpha = 10^\circ,20^\circ,...,90^\circ$; $\xi$ is the radial and $\phi_{\rm pc}$ the azimuthal coordinate in each case. The magnetic dipole axis $\mathbf{\mu}$ is situated at the origin, pointing outward normal to the plane of the page, in each case. The rotation axis $\mathbf{\Omega}$ is in the direction of $\phi_{\rm pc} = 0$, while the leading (trailing) edge of the pulse profile originates on B-field lines with footpoints around $\phi_{\rm pc}\sim90^\circ$ ($\phi_{\rm pc}\sim270^\circ$). The $\eta_{\rm c}$-solutions get progressively smaller for $\phi_{\rm pc}\sim180^\circ$, and for large $\alpha$, until no solution is found which satisfies our solution matching criteria (denoted by zero values or no values at all on the plots above). We ignore the emission from those particular field lines. We expect the $\eta_{\rm c}$-distribution to reflect the symmetry of the $\cos\phi_{\rm pc}$ function which is found in $E_{||}$; the small irregularities stem from the fact that we used interpolation on a non-uniform $(\xi,\phi_{\rm pc})$-grid when preparing the contour plots. (See online version for color plots.)\label{fig:eta_contour}}
\end{figure}

For illustration, Figure~\ref{fig:eta_contour} shows contour plots of $\eta_{\rm c}\sim1-6$ for different $\alpha$, $\xi,$ and $\phi_{\rm pc}$, and for $P=5$~ms; $\xi$ is the `radial' and $\phi_{\rm pc}$ the azimuthal coordinate for these polar plots. From these plots, one may infer that the ``oscillatory solutions'' are encountered when $\phi_{\rm pc}\sim180^\circ$, and for large $\alpha$ (which is where the second term of $E_{||}^{(2)}$ becomes negative and dominates the first positive term inside the square brackets of Eq.~[\ref{eq:E2}]). The $\eta_{\rm c}$-solutions become progressively smaller for these cases, until no solution is found which satisfies the above criteria; we ignore emission from those particular field lines.

\begin{figure}
\epsscale{0.8}
\plotone{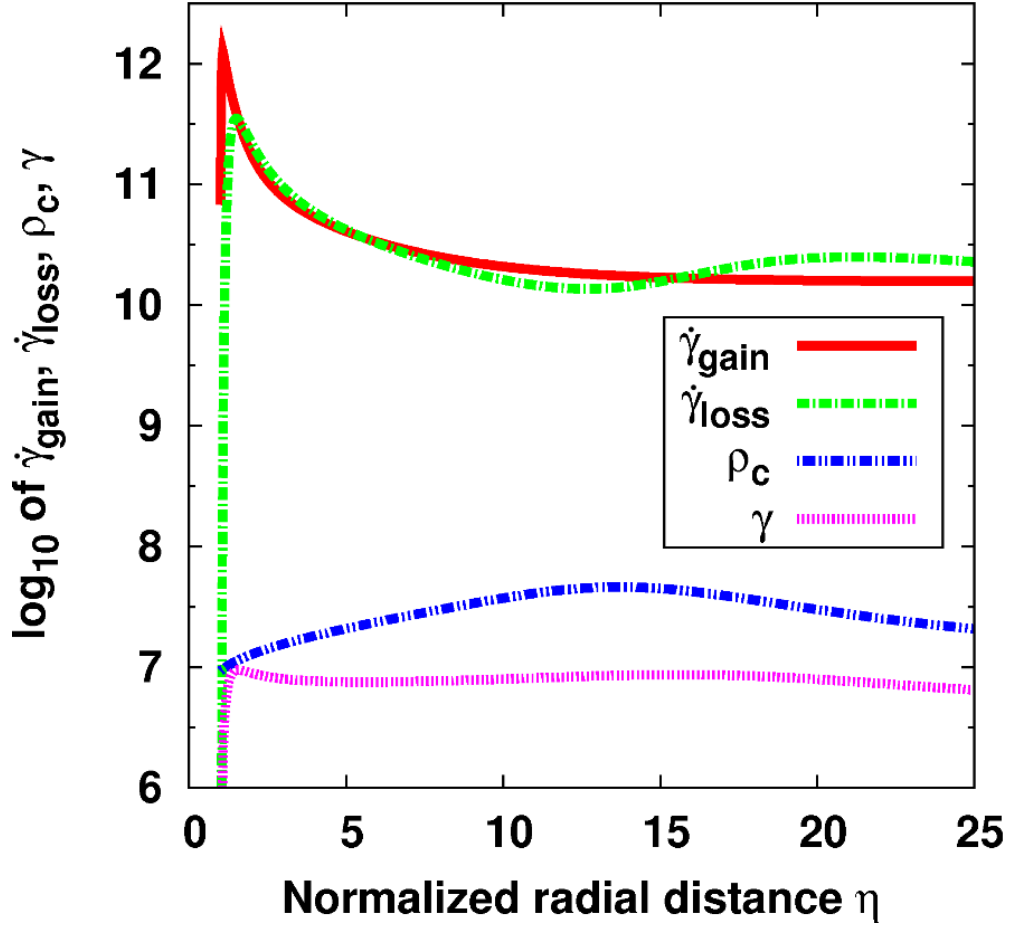}
\caption{The $\log_{10}$ of gain (acceleration) rate $\dot{\gamma}_{\rm gain}$ (solid line), loss rate $\dot{\gamma}_{\rm loss}$ (dash-dotted line), curvature radius $\rho_c$ (dash-dot-dotted line), and the Lorentz factor $\gamma$ (short-dashed line) as function of normalized radial distance $\eta$. We used $\phi_{\rm pc}=360^\circ$, $\xi=0.7$, $\alpha=40^\circ$, and $P=5$~ms in this plot.\label{fig:E1}}
\end{figure}

\begin{figure}
\epsscale{0.8}
\plotone{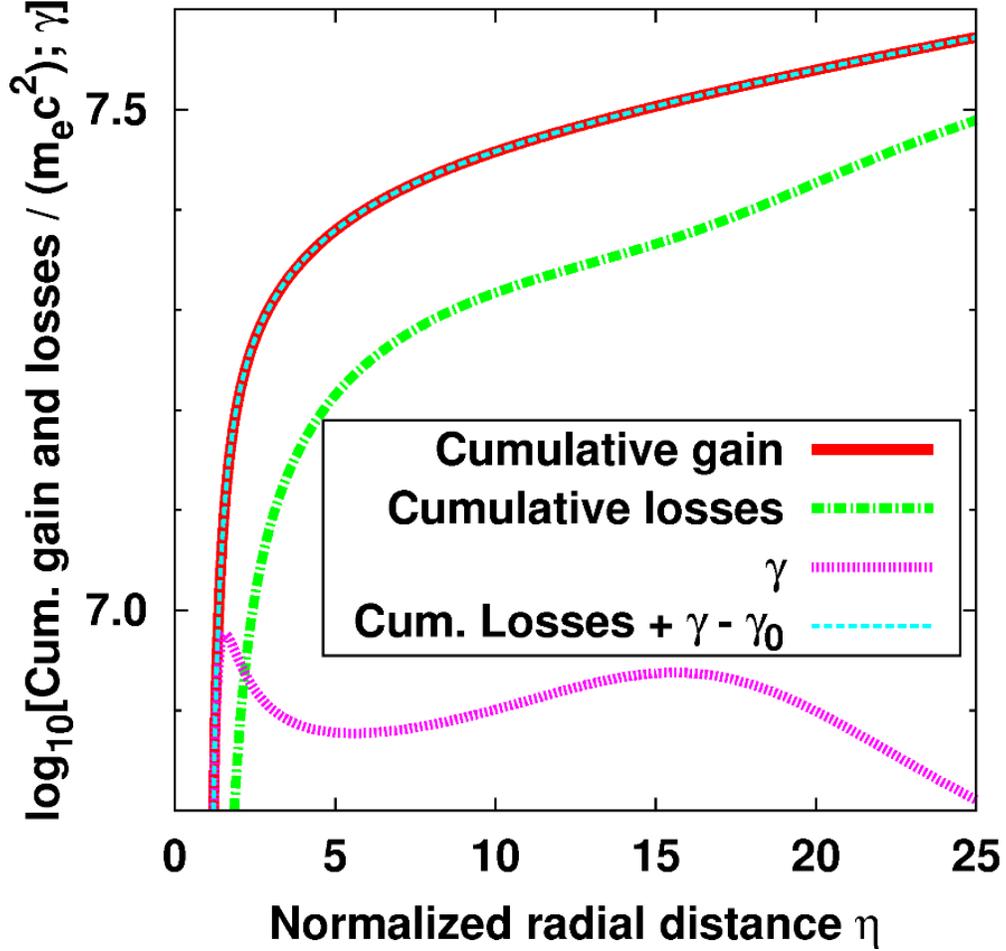}
\caption{The $\log_{10}$ of cumulative energy gain ($\int^\eta_{\eta=1} d\gamma_{\rm gain}$; solid line), cumulative energy losses ($\int^\eta_{\eta=1} d\gamma_{\rm loss}$; dash-dotted line), Lorentz factor $\gamma$ (short-dashed line), and the sum of the cumulative losses and acquired particle energy ($\int^\eta_{\eta=1} d\gamma_{\rm loss}+\gamma(\eta)-\gamma_0$; thin dashed gray / cyan line) in units of $m_ec^2$ vs.\ $\eta$. The latter sum and the cumulative gain coincide (within $\sim0.3\%$ for the $\eta$-range shown), pointing to energy balance, i.e.\ electric potential energy being converted into gamma-radiation and particle kinetic energy. We used $\phi_{\rm pc}=360^\circ$, $\xi=0.7$, $\alpha=40^\circ$, and $P=5$~ms for this plot.\label{fig:E2}}
\end{figure}

We tested our full solution of $E_{||}$, which incorporates $E_{||}^{(1)}$ through $E_{||}^{(3)}$, for conservation of energy when solving the transport equation (Eq.~[\ref{eq:transport}]) for relativistic electron primaries. Figure~\ref{fig:E1} indicates the $\log_{10}$ of acceleration rate $\dot{\gamma}_{\rm gain}=\dot{E}_{\rm e,gain}/m_ec^2$, loss rate $\dot{\gamma}_{\rm loss}=\dot{\gamma}_{_{\rm CR}}$, curvature radius $\rho_c$, and the Lorentz factor $\gamma$ as functions of distance. Although we did not find perfect radiation reaction where the acceleration and loss terms are equal in magnitude (similar to the findings of \citet{Venter_phd}), integration of these terms along different B-field lines yielded energy balance (i.e., conversion of electric potential energy into gamma-radiation and particle kinetic energy) for each integration step of the particle trajectory. An example of this is shown in Figure~\ref{fig:E2}, where the graph of the cumulative energy gain ($\int_{\eta = 1}^\eta d\gamma_{\rm gain}$) coincides with that of the sum of the cumulative energy losses and the acquired particle energy ($\int_{\eta=1}^\eta d\gamma_{\rm loss}+\gamma(\eta)-\gamma_0$) for all $\eta$ (to within $\sim$0.3\%), with $\gamma_0 = \gamma(\eta=1)$ the initial Lorentz factor at the stellar surface. We used $\gamma_0=100$, but the calculation is quite insensitve to this assumption, as $\gamma$ quickly reaches values of $\sim10^7$ (Figure~\ref{fig:E1}). The quantities in Figure~\ref{fig:E2} are plotted in units of $m_ec^2$.

\subsection{Generation of Phaseplots}
\label{sec:Phaseplots}
In the case of the PSPC model, we normalize the particle outflow along each B-line according to
\begin{equation}
	d\dot{N}(\xi,\phi_{\rm pc}) = -\frac{\rho_e(\eta=1,\xi,\phi_{\rm pc})}{e}dS\beta_0c,\label{eq:Ndot}
\end{equation}
with $d\dot{N}$ the number of particles leaving a surface patch $dS$ per unit time with initial speed $\beta_0c$ and $\rho_e$ is the charge density given by Eq.~(12) of \citet{HM98}.
The latter is equal to the GR equivalent of the Goldreich-Julian charge density at the NS surface. The expression in Eq.~(\ref{eq:Ndot}) is similar to the classical Goldreich-Julian expressions used by \citet{Story07}:
\begin{eqnarray}
	\dot{N}_{\rm GJ} & = & 1.3\times10^{30}B_{12}P^{-2}\quad{\rm particles\,\,s,}^{-1}\\
	\dot{n}_{\rm GJ} & = & \frac{\dot{N}_{\rm GJ}}{2\pi\left(1-\cos\theta_{\rm pc}\right)},
\end{eqnarray}
with $\dot{N}_{\rm GJ}$ the total number of particles injected from the PC per unit time, $B_{12}\equiv B_0/10^{12}$~G, $\dot{n}_{\rm GJ}$ the injected particle flux, and $d\dot{n}_{\rm GJ}\equiv\dot{n}_{\rm GJ}dS$ analogous to the GR quantity $d\dot{N}$ defined in Eq.~(\ref{eq:Ndot}). While the classical injection rate $d\dot{n}_{\rm GJ}$ is constant across the PC, the GR expression we use has both a $\xi$- and $\phi_{\rm pc}$-dependence. (Even though $d\dot{N}$ varies across the PC, we assume that it stays constant along B-lines, i.e.\ that it has no $\eta$-dependence.) 

We have distributed primary electrons uniformly across the PC using a constant step length $dl_{\rm ovc}$ along all rings between consecutive electron positions, so that there are generally less electrons per ring for the inner rings than for the outer ones. Because of this uniform distribution, we could approximate the area of the surface elements using
\begin{equation}
	dS \approx \frac{\pi R_{\rm PC}^2}{N_{\rm e,tot}},
\end{equation}
with $N_{\rm e,tot}$ the total number of electrons positioned on the PC surface (depending on grid size of the mesh into which the PC area was divided). These electron positions coincide with B-line footpoints on the stellar surface. We next followed the motion of electron primaries along these lines (Section~\ref{sec:Transport}), collecting HE radiation and binning as described below.

The instantaneous CR power spectrum is given by \citep{Jackson75,Harding81,Daugherty82,Story07}
\begin{equation}
\left(\frac{dP}{dE}\right)_{\rm CR}=\sqrt{3}\alpha_{\rm fine}\gamma\left(\frac{c}{2\pi\rho_c}\right)\kappa\left(\frac{\epsilon_\gamma}{\epsilon_{\rm CR}}\right),
\end{equation}
with 
\begin{equation}
\frac{\epsilon_{\rm CR}}{m_ec^2} = \frac{3\lambdabar_c\gamma^3}{2\rho_{\rm c}} = \frac{3\hbar c\gamma^3}{2m_ec^2\rho_{\rm c}}
\end{equation}
the critical energy, $\lambdabar_c=\hbar/(m_ec)$ the Compton wavelength, $\alpha_{\rm fine}$ the fine-structure constant, and \citep{Erber66}
\begin{equation}
\kappa(x) \equiv x\int_{x}^{\infty}K_{5/3}(x^{\prime})dx^{\prime} \approx \left\{
    \begin{array}{ll}
      2.149\,x^{1/3} & x \ll 1\\
      1.253\,x^{1/2}e^{-x} & x \gg 1,\label{eq:K53}
    \end{array}
    \right.
\end{equation}
with $K_{5/3}$ the modified Bessel function of order $5/3$. We calculate the number of CR photons radiated per unit time by the primaries in a spatial step $ds_{\rm IOF}$ (as measured along the B-field line in the IOF), in an energy bin of width $dE = E_2-E_1$, using
\begin{equation}
	d\dot{n}_{\rm \gamma,CR} = \frac{\dot{\gamma}_{_{\rm CR}}W}{\overline{E}_{\rm bin}}×\frac{ds_{\rm IOF}}{c}×d\dot{N},\label{eq:ncr}
\end{equation}
with
\begin{equation}
	\overline{E}_{\rm bin} = \frac{1}{2m_ec^2}\left(E_1+E_2\right),
\end{equation}
and
\begin{equation}
	W = \frac{\int_{E_1}^{E_2}\kappa(x)\,dx}{\int_{E_0}^\infty \kappa(x)\,dx},
\end{equation}
with $E_0\ll1$. The expression in Eq.~(\ref{eq:ncr}) gives the number of photons radiated per primary per unit time with an energy $\sim\overline{E}_{\rm bin}$ (i.e., the ratio of power radiated per primary in a particular energy bin to average bin energy, in $m_ec^2$ units) multiplied by a time step $ds_{\rm IOF}/c$, multiplied by the number of primaries passing per unit time $d\dot{N}$. (The `weighting factor' $W$ therefore scales the total power to the power radiated in the particular energy bin.) We ignore field lines with $d\dot{N}<0$.

For all the other geometric models, we assume constant emissivity per unit length, i.e.\ $d\dot{n}_{\rm \gamma,CR}\propto ds_{\rm IOF}$.

We lastly accumulate $d\dot{n}_{\rm \gamma,CR}$ in $(\zeta,\phi)$-bins (after applying the SR effects described in Section~\ref{sec:B}), and divide by the solid angle subtended by each phaseplot bin, $d\Omega = (\cos\zeta-\cos(\zeta+d\zeta))d\phi \approx\sin\zeta d\zeta d\phi$, to make up the final phaseplot.

\subsection{Radio Beam Model}
We model the radio emission beam using an empirical cone model that has been developed over the years through detailed study of pulse morphology and polarization characteristics of the average-pulse profile. The average-pulse profiles are quite stable over long timescales and typically show a variety of shapes, ranging from a single peak to as many as five separate peaks. The emission is also highly polarized, and displays changes in polarization position angle across the profile that often matches the position angle swing expected for a sweep across the open field lines near the magnetic poles in the Rotating Vector Model \citep{RC69}. 

Rankin's \citep{Rankin93} study of pulse morphology concluded that pulsar radio emission can be characterized as having a core beam centered on the magnetic axis and one or more hollow cone beams also centered on the magnetic axis surrounding the core. Although Rankin's model assumes that emission fills the core and cone beams, other studies \citep[e.g.,][]{Lyne88} conclude that emission is patchy and only partially fills the core and cone beam patterns. 

The particular description we adopt is from \citet{Gonthier04} and is based on work of Arzoumanian et al.\ \citep{ACC02}, who fit average-pulse profiles of a small collection of pulsars at 400~MHz to a core and single cone beam model based on the work of Rankin. The flux from the conal component seen at angle $\theta$ to the magnetic field axis (modified by \citet{Gonthier04} to include frequency dependence $\nu$) is 
\begin{equation}  \label{eq:Stheta}
S(\theta, \nu ) = F_{\rm cone} e^{ - (\theta  - \bar \theta )^2 /\omega _e^2 }.
\end{equation}
The annulus position and width of the cone beam are
\begin{equation} \label{eq:thetabar}
\bar \theta  =  (1.0-2.63\,\delta_w) \rho_{\rm cone},
\end{equation}
\begin{equation}  \label{eq:widann}
w_e = \delta_w \rho_{\rm cone},
\end{equation}
where $\delta_w = 0.18$ \citep{Harding07_Geminga}, and
\begin{equation}  \label{rhocone}
\rho_{\rm cone} = 1.24^{\circ}\,  r_{\rm KG}^{0.5}\, P^{- 0.5},
\end{equation}
with
\begin{equation}  \label{eq:rKG}
r_{\rm KG}  \approx 40\, \left({\dot P\over 10^{ - 15}{\rm s\,s^{-1}}}\right)^{0.07} P^{0.3} \nu_{\rm GHz}^{ - 0.26}
\end{equation}
the radio emission altitude in units of stellar radius \citep{Kijak03}, and $\nu_{\rm GHz}\equiv\nu/1$~GHz. (We do not assume a longitudinal extension of the radio emission region, but only use a single emission altitude.) According to Eq.~(\ref{eq:rKG}), the altitude of the conal radio emission is a weak function of $P$, but the emission occurs increasingly close to the light cylinder (at $R_{\rm LC}=c/\Omega$) as $P$ decreases (for more or less constant $\dot{P}$). For Crab-like periods, the conal emission occurs at altitudes of $10\% - 20\%$ of the light cylinder radius (and similar for typical MSP parameters of $P\sim$ a few milliseconds and $\dot{P}\approx10^{-20}$). For the current study, we are only interested in pulse shapes and phase shifts between the radio and gamma-ray pulses. We therefore use relative units for the cone beam luminosity.

\subsection{Flux Correction Factor}
\label{sec:fom}
It is very important to be able to scale from the observed (phase-averaged) energy flux $G_{\rm obs}$ to the all-sky luminosity, as this is used to define the gamma-ray radiation efficiency $\eta_\gamma$, a crucial quantity in characterizing the energetics of pulsar emission (see e.g.\ \citet{Abdo09_Cat}, where $\eta_\gamma\propto f_\Omega$). Such a flux correction factor ($f_\Omega$) is necessarily model-dependent, as any observer only sees a small part of the total radiation: that coming from a slice through the emission beam, determined by the line-of-sight $\zeta$.

\citet{Venter_phd} defined the total gamma-ray luminosity using
\begin{equation}
 L_\gamma = \Lambda d^2 G_{\rm obs},
\end{equation}
with $\Lambda=\varepsilon\overline{\Delta\Omega}^{\rm beam}/\beta^{\rm obs}$, $\varepsilon = \beta^{\rm obs}G^{\rm beam}/G^{\rm obs}$, $\beta^{\rm obs}$ the duty cycle, $\overline{\Delta\Omega}^{\rm beam}$ the average beaming angle, and $G^{\rm beam}$ the all-sky total energy flux. \citet{Watters09} used a similar definition
\begin{equation}
 L_\gamma = 4\pi f_\Omega d^2 G_{\rm obs},
\end{equation}
\begin{equation}
f_\Omega(\alpha,\zeta_E) = \frac{\int\!\!\!\int F_\gamma(\alpha,\zeta,\phi)\sin\zeta d\zeta d\phi}{2\int F_\gamma(\alpha,\zeta_E\phi)\,d\phi},\label{eq:fom}
\end{equation}
with $F_\gamma$ the photon flux per solid angle (`intensity'), and $\zeta_E$ the Earth line-of-sight, so that
\begin{equation}
 \Lambda \approx 4\pi f_\Omega,
\end{equation}
assuming similar distributions of gamma-ray photon and energy fluxes in $(\zeta,\phi)$-space (i.e.\ $F_{\gamma}(\zeta,\phi)/F_{\rm \gamma,tot}\approx G_{\gamma}(\zeta,\phi)/G_{\rm \gamma,tot}$). In Section~\ref{sec:Results}, we calculate $f_\Omega$ for different pulsar models, using Eq.~(\ref{eq:fom}).

\section{Light Curve Data}
\label{sec:LCs}
We compare the light curves generated with the different theoretical models (by making constant-$\zeta$ cuts through the respective phaseplots of gamma-ray and radio emission) to the light curves of the eight MSPs recently discovered by \textit{Fermi}-LAT \citep{Atwood09,Abdo09_MSP} in the right panels of Figures~\ref{fig:LC_com1} through~\ref{fig:LC_com4}.

The \textit{Fermi}-LAT light curves were produced by phase-folding LAT photons with energies above 100~MeV, recorded between 30 June 2008 and 15 March 2009. In order to reduce the contamination of the gamma-ray signal by the Galactic and extragalactic diffuse emission or nearby sources, and thereby maximize the signal-to-noise ratio, photons were selected in narrow regions of interest, with radii of $0.5^\circ$ to $1^\circ$ around the pulsar locations. The gamma-ray light curves seen by the LAT are shown in the right panels of Figures~\ref{fig:LC_com1} through~\ref{fig:LC_com4}, along with the radio profiles providing the absolute phase alignment. As the models predict different radio-to-gamma lags $\delta$, the phase alignment is crucial. The horizontal dashed lines indicate the background level estimated from a ring surrounding the pulsar.

It is important to note that the LAT angular resolution depends on the photon energy: the 68\% containment radius is $3.5^\circ$ at 100~MeV, and $0.6^\circ$ at 1~GeV \citep[see][]{Atwood09}. A consequence of the narrowly-chosen regions of interest is that a significant fraction of low-energy photons emitted by the pulsars are rejected. Therefore, the light curves shown in Figures~\ref{fig:LC_com1} through~\ref{fig:LC_com4} are biased toward energies above 1~GeV and may not reflect the actual profile shape obtained using larger regions of interest.

As the \textit{Fermi} mission continues, increased photon counts will allow the study of light curve shape as a function of energy in more detail. In fact, updated gamma-ray profiles based on energy-dependent angular cuts do not differ fundamentally from what is seen in Figures~\ref{fig:LC_com1} through~\ref{fig:LC_com4} \citep{Guillemot09}.

\section{Results}
\label{sec:Results}
Table~\ref{tab1} summarizes some of the properties of the MSPs discovered by \textit{Fermi}-LAT \citep{Abdo09_MSP}. All distances come from parallax measurements, except for those of PSR~J0218+4232 and PSR~J1614$-$2230 which are based on the NE2001 model (see \citet{Abdo09_MSP} for references). The values of $\dot{P}$ have been corrected for the Shklovskii effect \citep{Shklovskii70}.

\begin{figure}
\epsscale{0.8}
\plotone{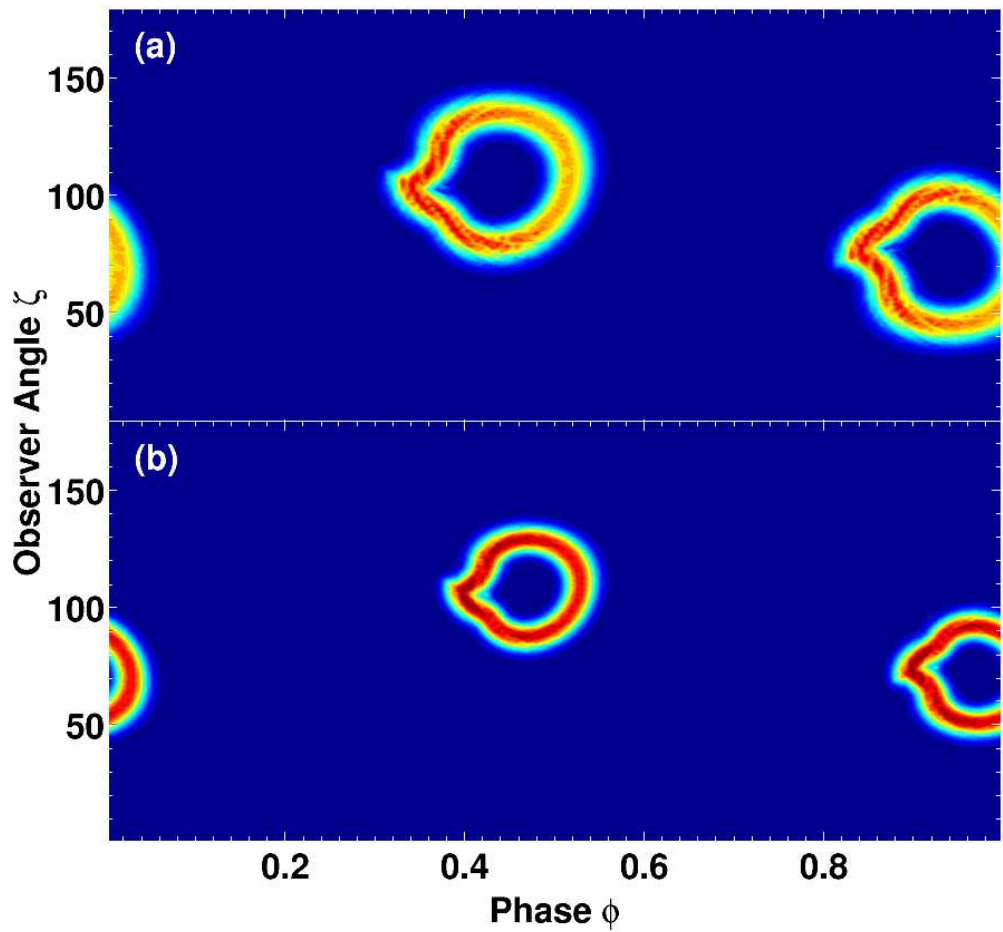}
\caption{Example phaseplots of the radio conal beam, for $\alpha=70^\circ$ and at a frequency of 1.4~GHz. Panel~(a) is for $P = 2$~ms, and panel~(b) for $P = 5$~ms. Beam and annulus widths become increasingly larger for shorter periods. Increased intensity for the leading part is due to bunching of the B-field lines around the notch. (Note that in this and following phaseplots, the color scales are not the same for the different panels, but are chosen to show the most details for each case.)\label{fig:P_rad}}
\end{figure}

The radio beam may be quite large in the case of MSPs. Figure~\ref{fig:P_rad} shows examples of phaseplots of the radio conal beam for $\alpha=70^\circ$. The top panel is for $P = 2$~ms, and the bottom one for $P = 5$~ms. The conal beam's total size and annular width become increasingly larger for shorter periods, scaling as $P^{-0.35}$. The notch, a feature of the retarded magnetic field solution \citep{Dyks04_B}, is apparent as well as increased intensity for the leading part, which is due to bunching of the B-field lines around the notch. (In the online version, plots are shown in color.)

\begin{figure}
\epsscale{0.8}
\plotone{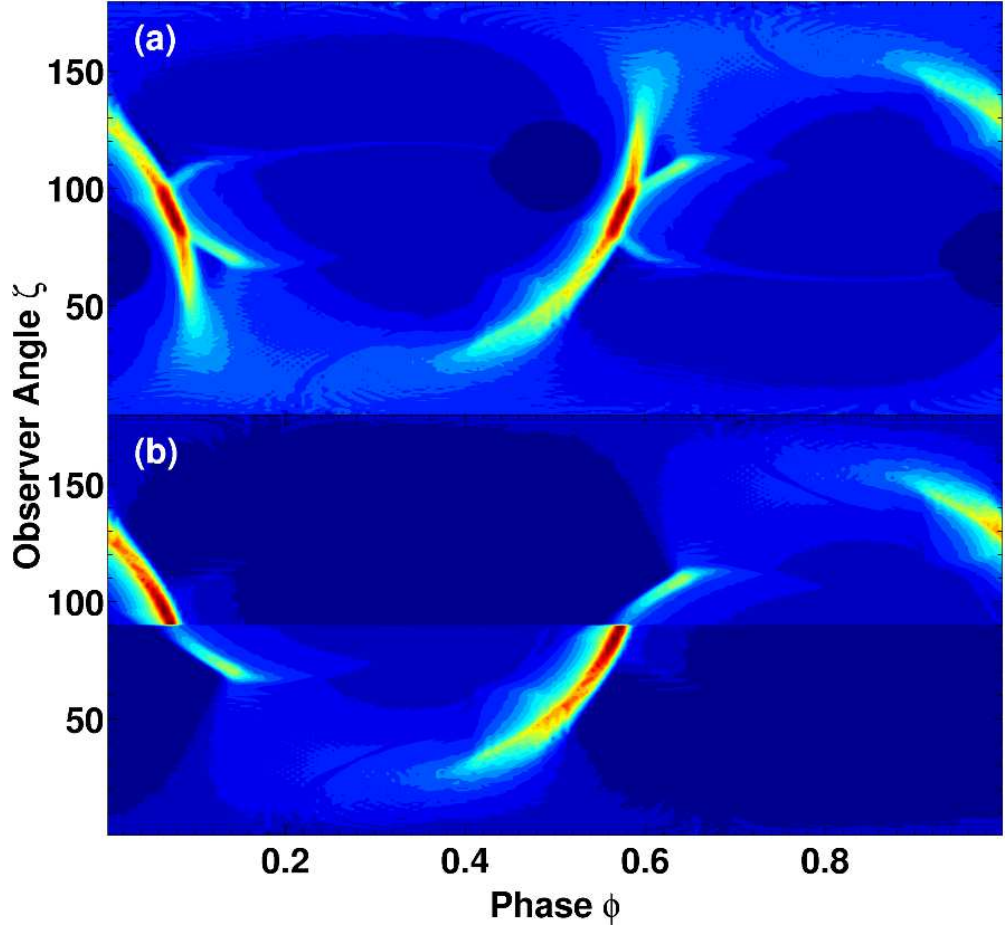}
\caption{Example phaseplots for the TPC2 and OG1 models (panel~(a) and~(b) respectively), for $\alpha=70^\circ$ and $P=5$~ms. In contrast to the TPC models, no emission originates below the null charge surface in the OG model, in which case an observer can only see emission from one magnetic pole.\label{fig:P1}}
\end{figure}

\begin{figure}
\epsscale{0.8}
\plotone{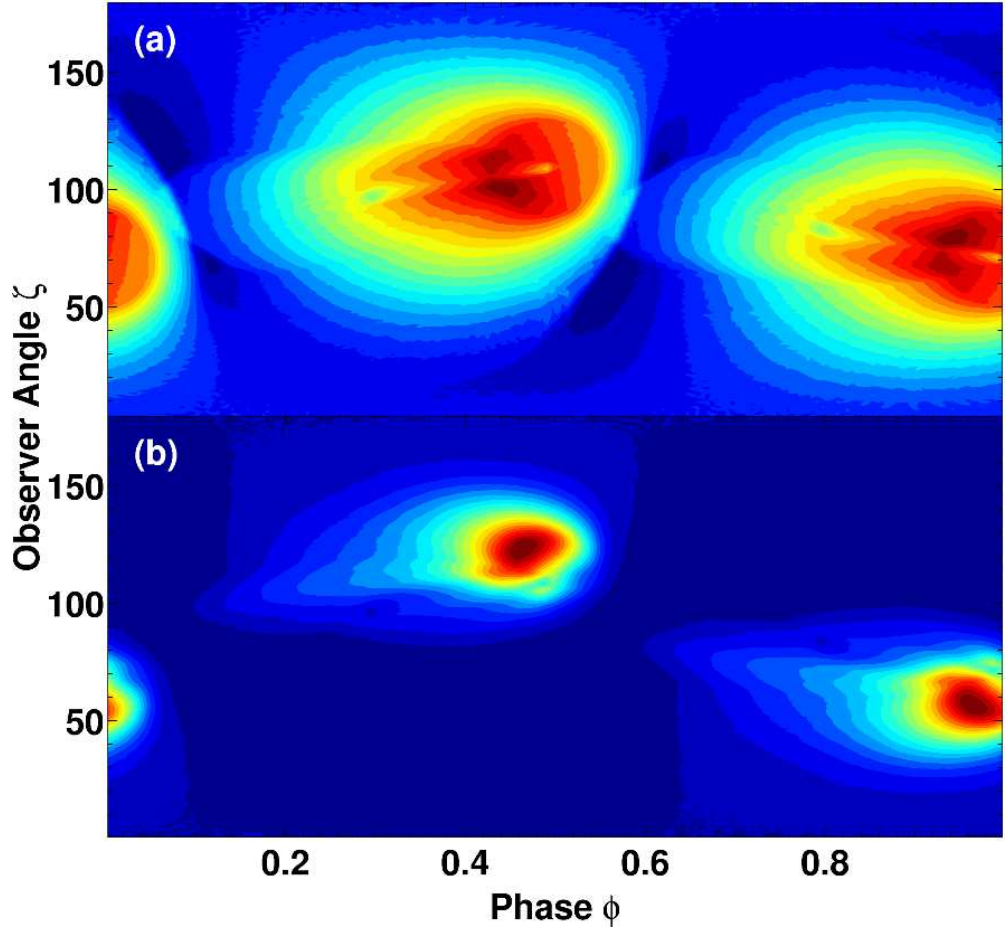}
\caption{Example phaseplots for the PC1 and PC2 models (panel~(a) and~(b) respectively), for $\alpha=70^\circ$ and $P=5$~ms.\label{fig:P2}}
\end{figure}

Differences of the geometric TPC, OG, PC, and also the PSPC models are graphically presented in Figures~\ref{fig:P1} and~\ref{fig:P2} (the geometric PC models do not provide particularly good fits to the observed light curves, and we will therefore not concentrate on their detailed properties in what follows). Figure~\ref{fig:P1} shows example phaseplots for TPC (top panel) and OG (bottom panel) models for $\alpha=70^\circ$. For emission tangent to trailing field lines, relativistic effects of aberration and time-of-flight delays cause phase shifts that nearly cancel those due to the curvature of the B-field, leading to accumulation of emission around narrow phase bands. This yields caustic structures around $\sim0.0-0.1$ and $\sim0.4-0.6$ in phase. (The observer phase $\phi$ is defined to be zero where the observer crosses the meridional plane which contains both $\mathbf{\Omega}$ and the magnetic dipole axis $\mathbf{\mu}$.) Emission is assumed to be symmetric for both magnetic poles. In the OG model, no emission originates below the null charge surface (note that the null charge surface is at $\zeta=90^\circ$ in these plots), so that an observer can only see emission from one magnetic pole, in contrast to the SG / TPC models where an observer sees emission from both poles. Therefore, by comparing the two panels of Figure~\ref{fig:P1}, one can infer which part of the caustics originate at low emission altitudes (present only in TPC models), and which part at high altitudes (present in both TPC and OG models). The dark circular structures at phase 0 and 0.5 in the TPC-case are the PC surfaces from opposite poles. They are significantly larger for MSPs than for younger pulsars, since their size scales with $P^{-1/2}$.

Figure~\ref{fig:P2} shows example phaseplots for the constant-emissivity PC case (top panel), and a PSPC model (bottom panel) including the full GR E-field, for $\alpha=70^\circ$. The difference in shape of the emission regions associated with the magnetic axes in the latter case reflects the dependencies of the E-field on spatial parameters. The emission regions are also much smaller (implying correspondingly smaller gamma-ray peak widths), as the E-field decreases with altitude before reaching a constant value (Figure~\ref{fig:etac}). 

\begin{figure}
\epsscale{0.8}
\plotone{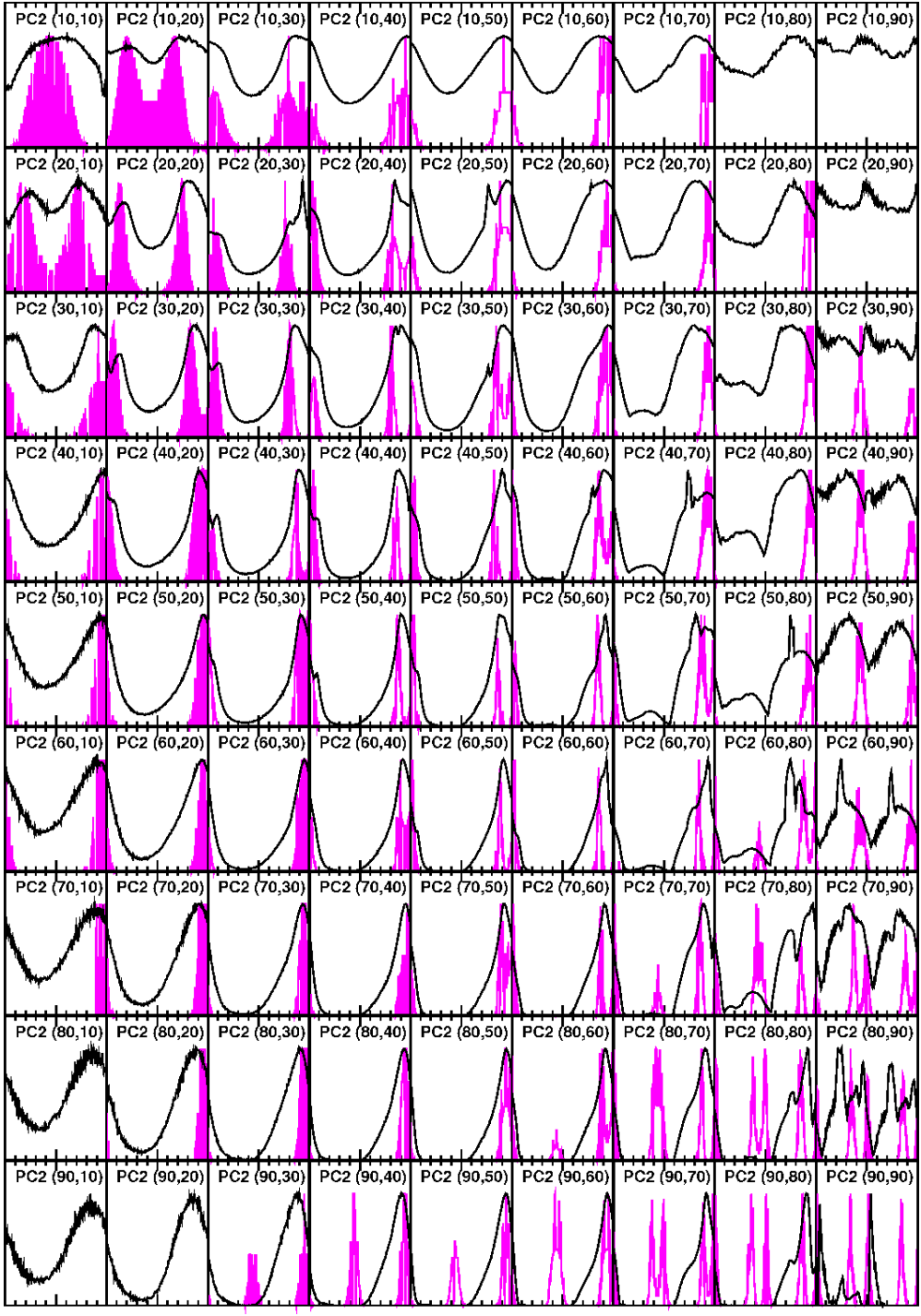}
\caption{Sample light curves (black: gamma-ray; gray / magenta: radio at 1.4~GHz) for the PC2 model, with $P=2$~ms. The observer angle $\zeta$ changes along the columns, and the inclination angle $\alpha$ along the rows. All pulse shape maxima are normalized to unity, and the phase range goes from $\phi=0 - 1$ in each case (and similar for subsequent figures).\label{fig:LC1}}
\end{figure}

\begin{figure}
\epsscale{0.8}
\plotone{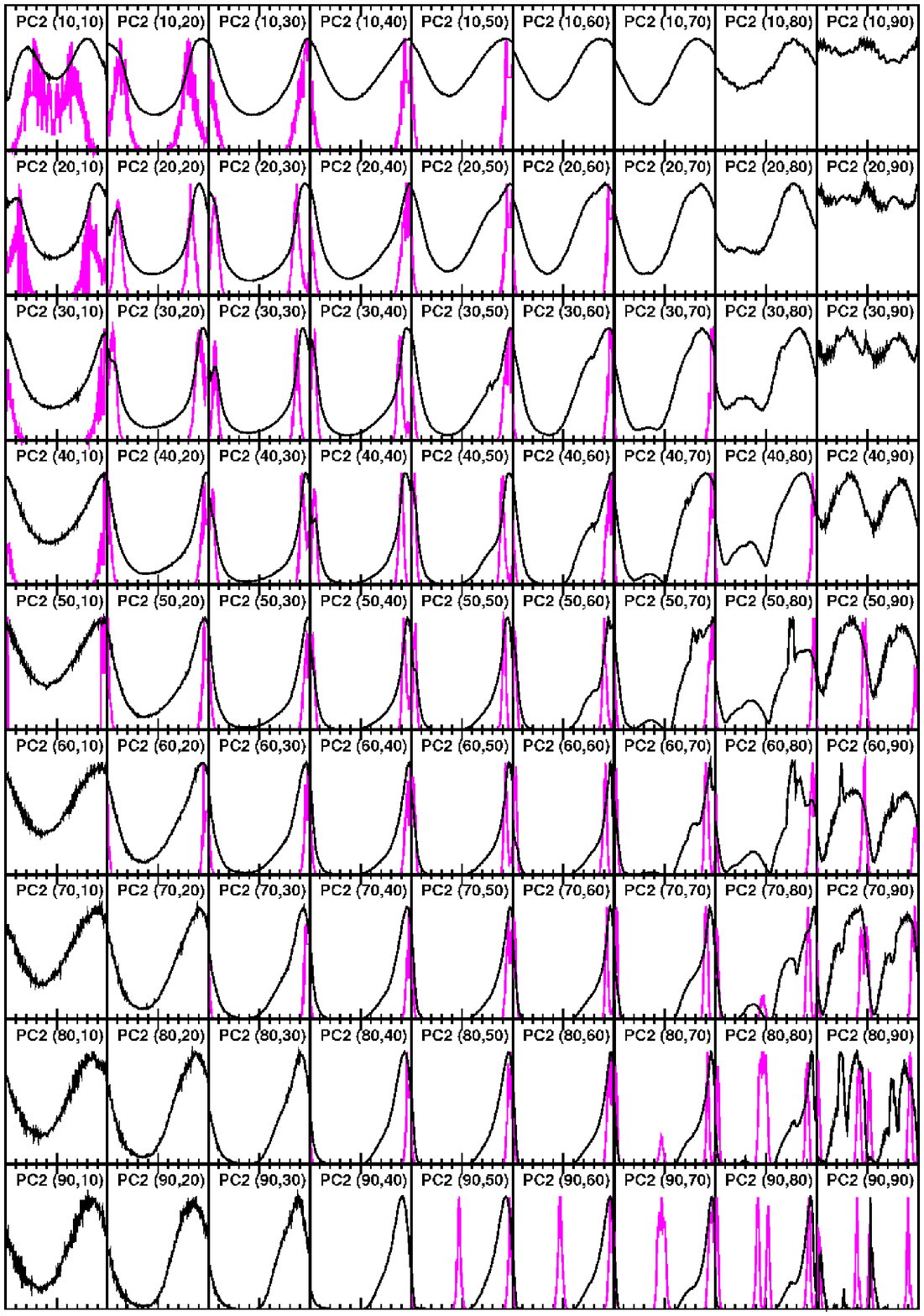}
\caption{Sample light curves for a PC2 model with $P=5$~ms.\label{fig:LC2}}
\end{figure}

\begin{figure}
\epsscale{0.8}
\plotone{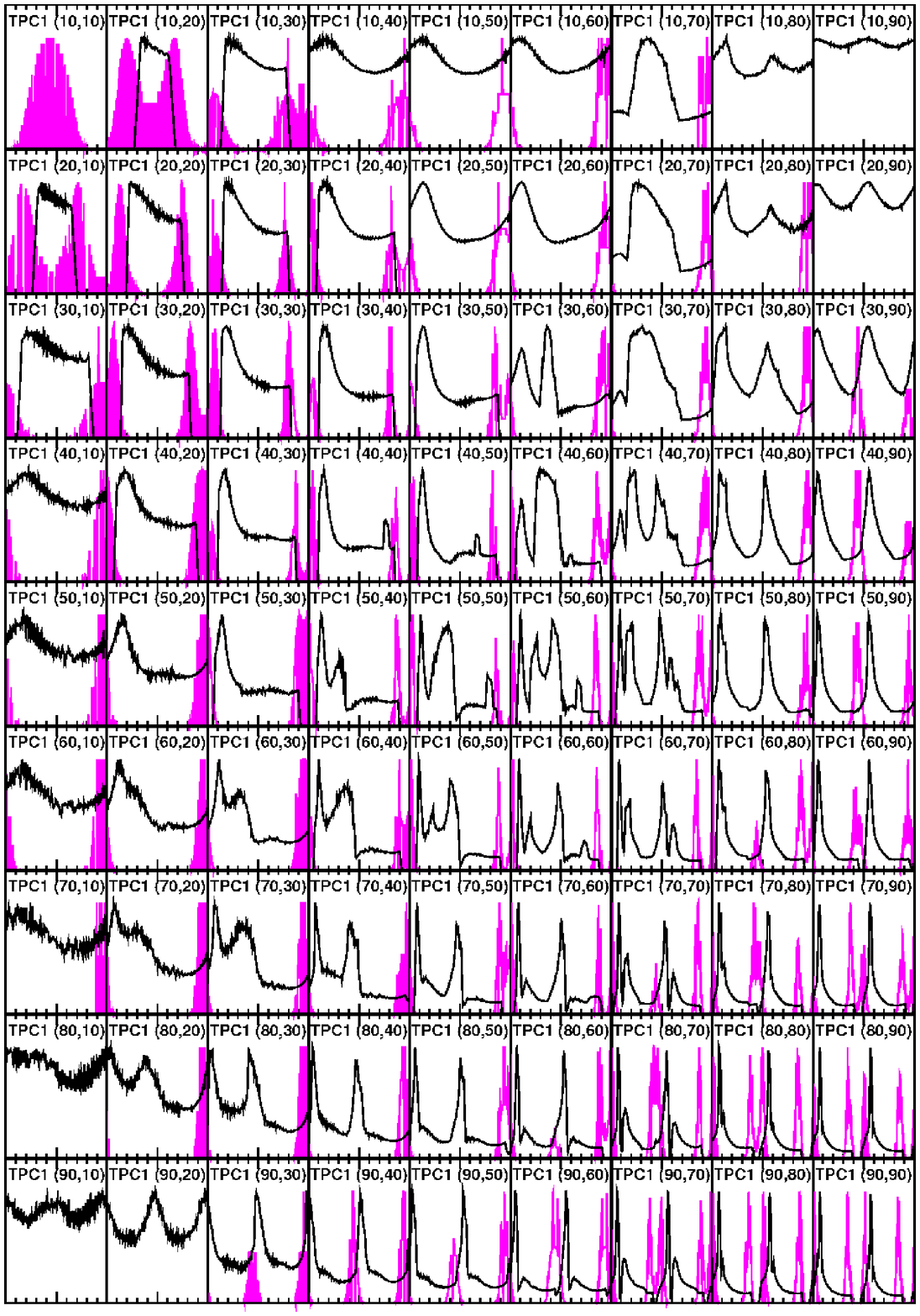}
\caption{Sample light curves for a TPC1 model with $P=2$~ms.\label{fig:LC3}}
\end{figure}

\begin{figure}
\epsscale{0.8}
\plotone{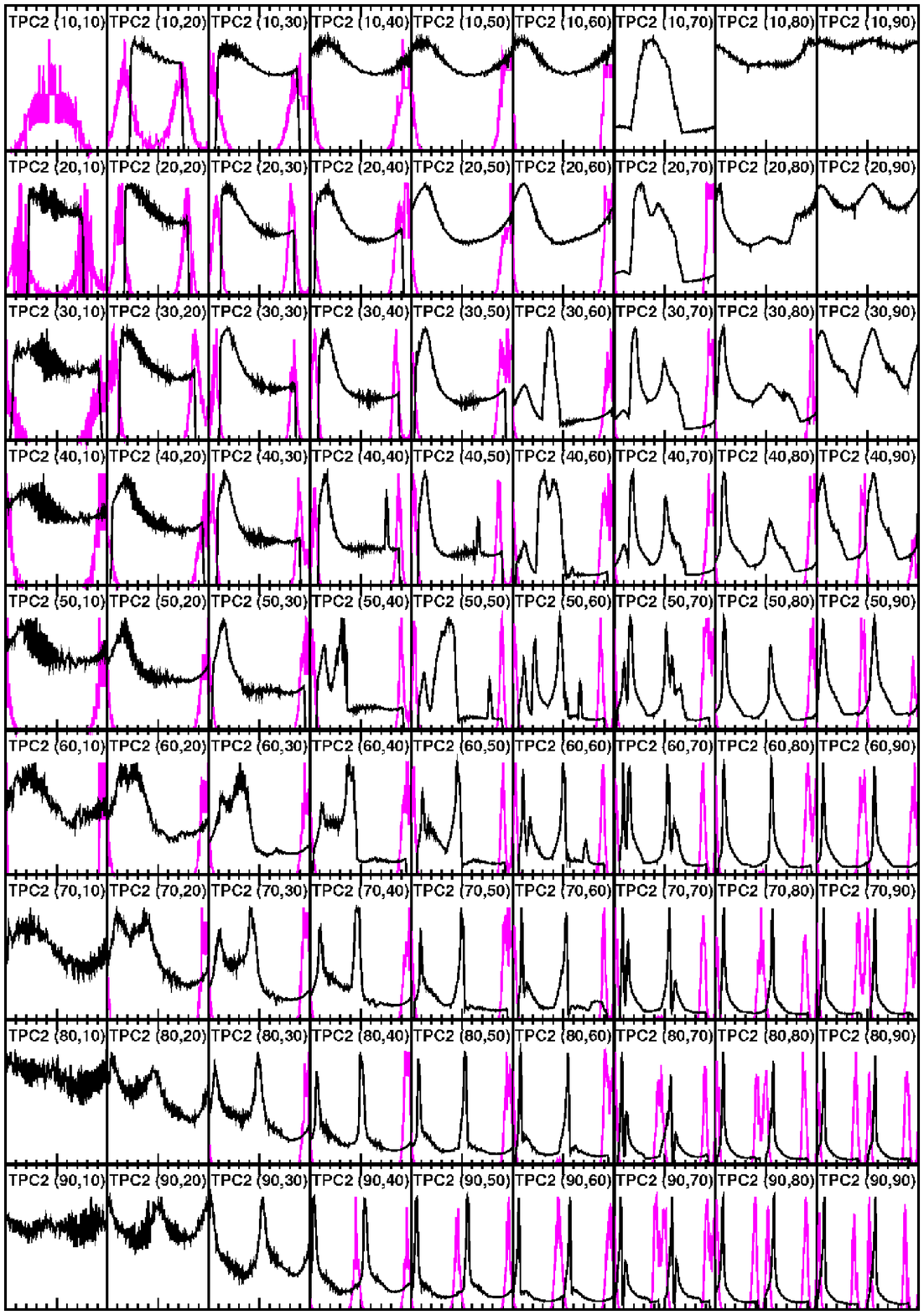}
\caption{Sample light curves for a TPC2 model with $P=3$~ms.\label{fig:LC4}}
\end{figure}

\begin{figure}
\epsscale{0.8}
\plotone{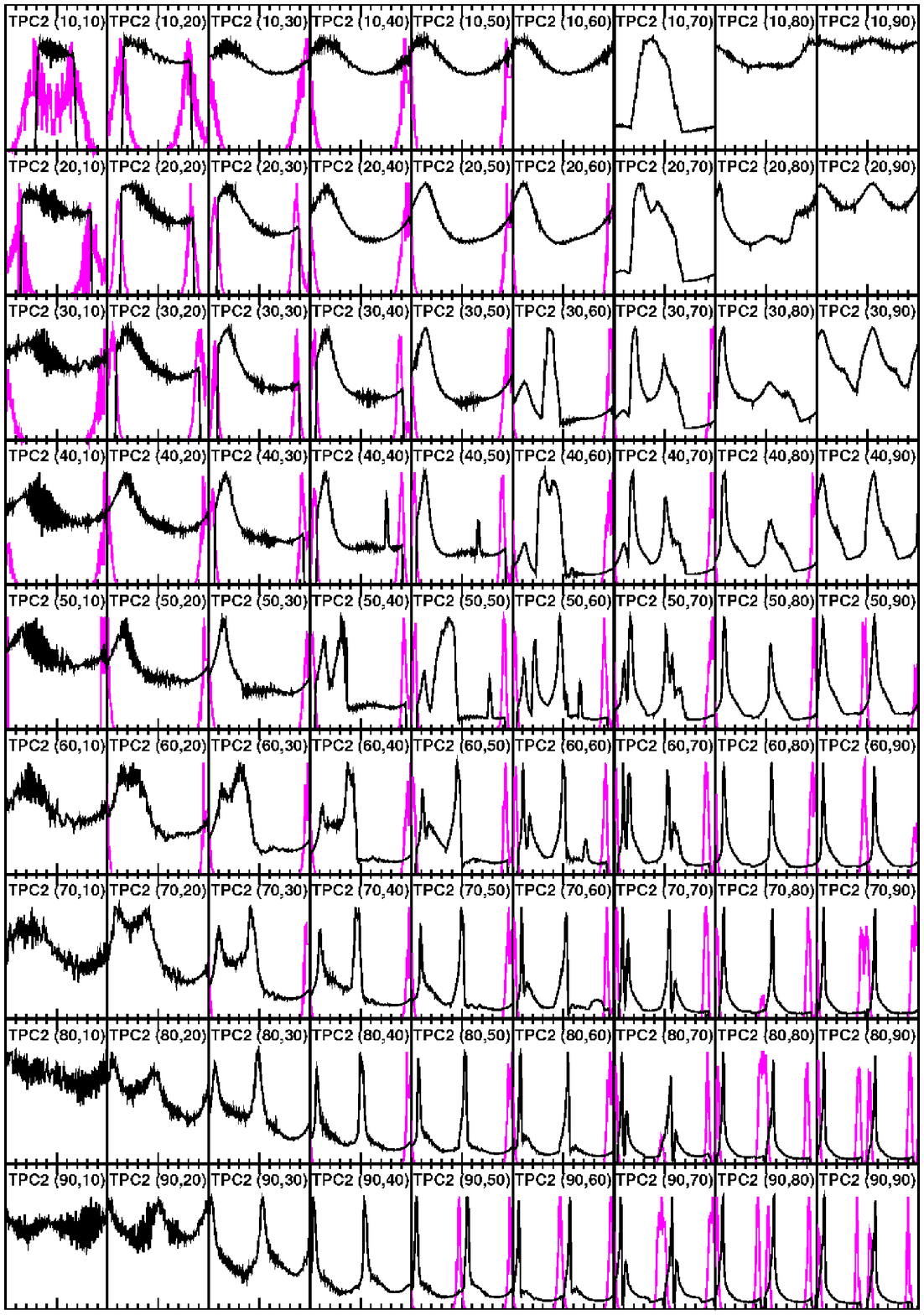}
\caption{Sample light curves for a TPC2 model with $P=5$~ms.\label{fig:LC5}}
\end{figure}

\begin{figure}
\epsscale{0.8}
\plotone{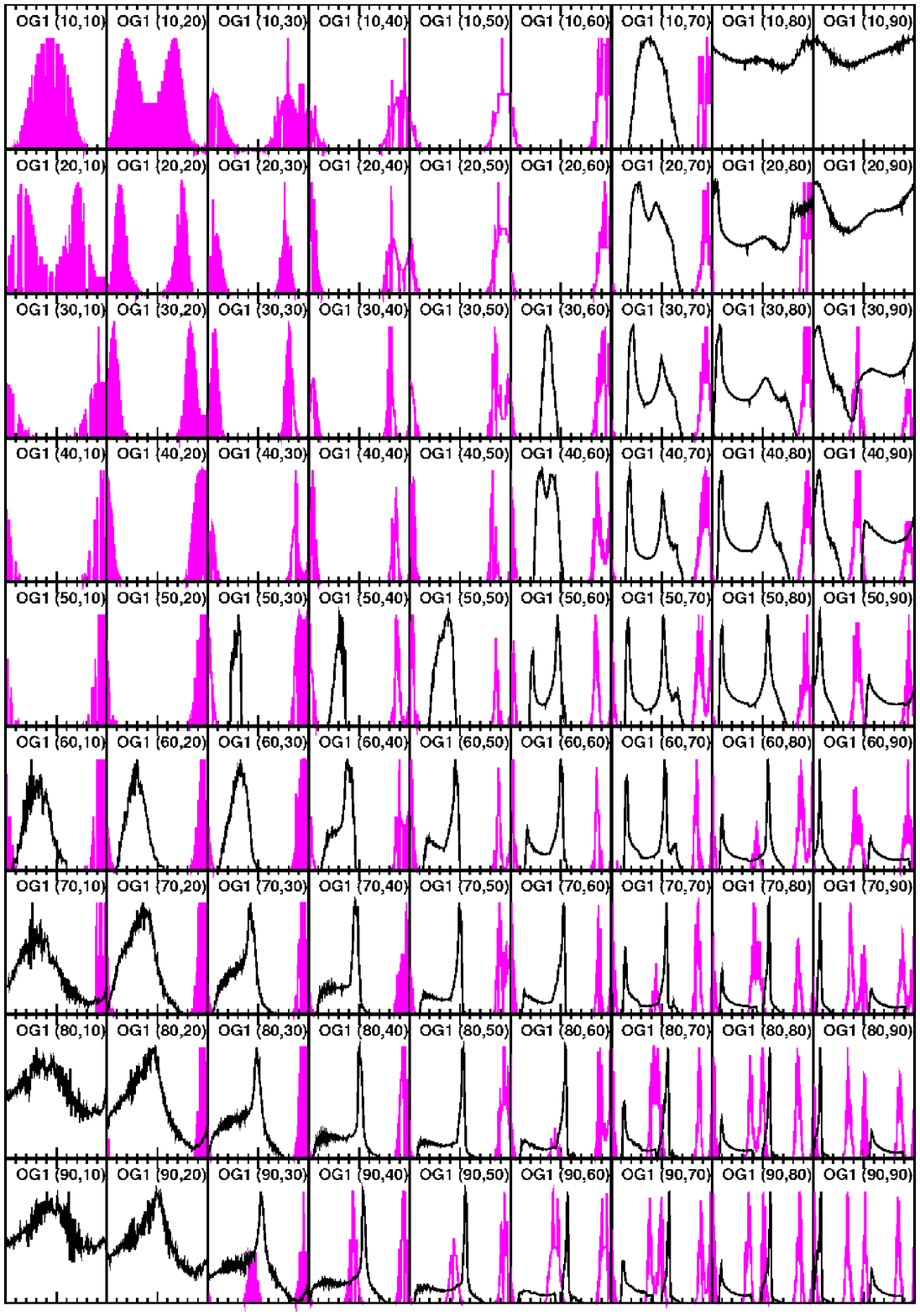}
\caption{Sample light curves for an OG1 model with $P=2$~ms.\label{fig:LC6}}
\end{figure}

\begin{figure}
\epsscale{0.8}
\plotone{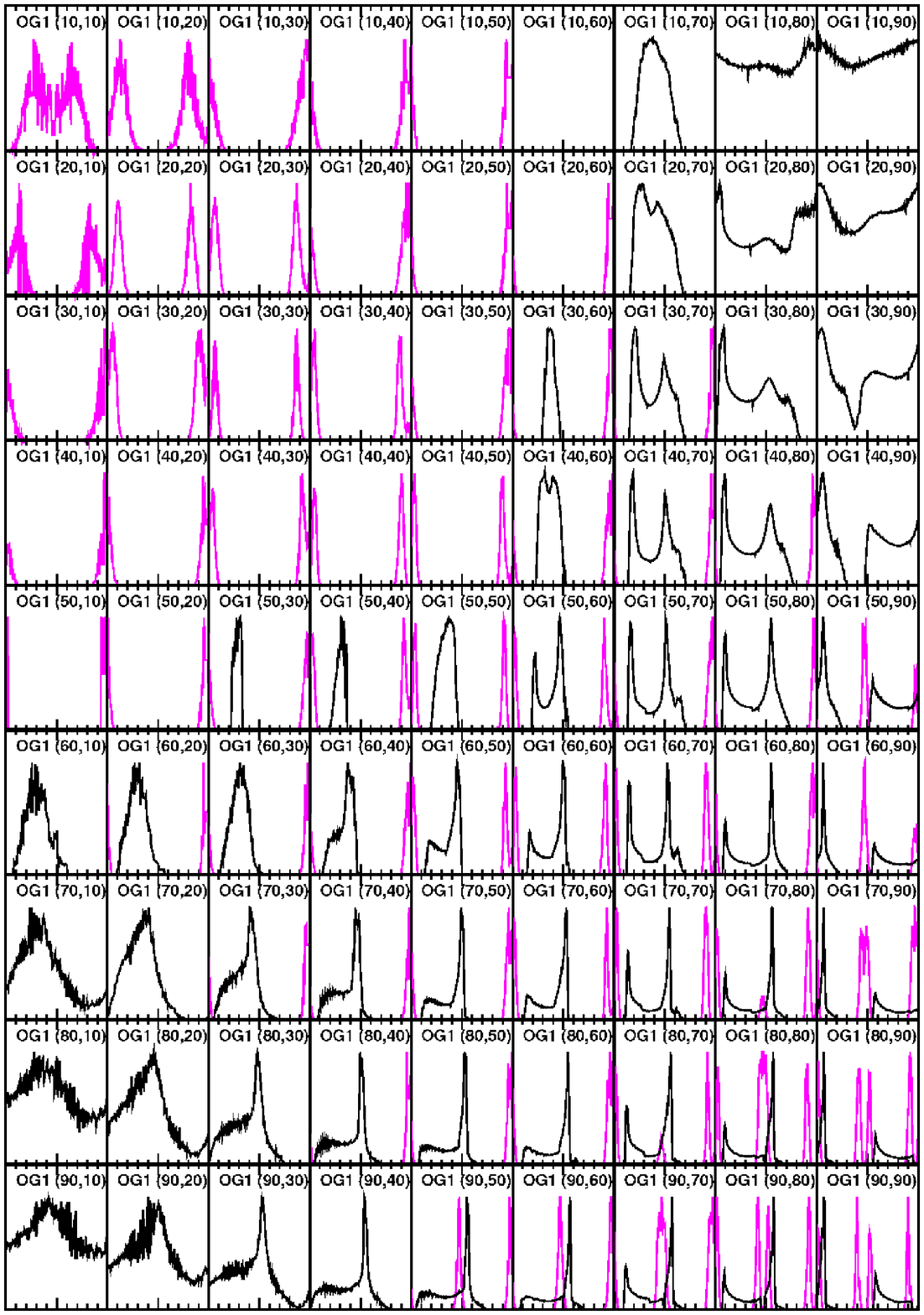}
\caption{Sample light curves for an OG1 model with $P = 5$~ms.\label{fig:LC7}}
\end{figure}

In order to fit the \textit{Fermi}-LAT and radio light curves (and to compare different model predictions; see Section~\ref{sec:LCs}), we generated a large number of light curves for each of the different pulsar models, and for nearly the full range of inclination and observer angles ($\alpha = \zeta = 5^\circ-90^\circ$, in $5^\circ$ intervals); also for $P=2$, 3, and~5~ms, and for different gap widths (Section~\ref{sec:B}). (Although the phaseplots are usually very similar for different $P$ in the case of younger pulsars, the PC size is significantly larger in the MSP case, and may impact light curves derived from the phaseplots.) Some example light curves generated using different phaseplots are shown in Figure~\ref{fig:LC1} through~\ref{fig:LC7} (see Table~\ref{tab3} for explanation of the model abbreviations used). Each panel shows the light curves (black: gamma-ray, gray / magenta: radio) corresponding to different ($\alpha$,$\zeta$)-combinations, with the normalized phase $\phi=0-1$ in each case. Note that all profiles have been renormalized, since we were primarily interested in pulse shape (and radio-to-gamma phase lag). This has the effect of boosting low-level emission, leading to noisy profiles in some cases (e.g., the first column of Figure~\ref{fig:LC1}). Details as to the model, and chosen period $P$, are given in the captions of these Figures.

The PSPC (and PC) model have mostly single-peaked gamma-ray profiles which are roughly in phase with the radio (when there is only a single radio peak), and the profiles become larger when $P$ decreases (especially the radio). The radio profile may exhibit zero, one or two peaks, depending on where the observer's line-of-sight intersects with the radio cone. In significantly off-beam geometries (large impact angle $\beta = \zeta-\alpha$), one therefore only sees gamma-ray radiation (i.e., missing the radio cone), in accordance with expectations that gamma-ray beams are larger than their radio counterparts. This is the standard way of explaining the phenomenon of `radio-quiet' pulsars \citep[e.g.,][]{Abdo09_Cat}. 
Double-peaked radio profiles occur for both large $\alpha$ and $\zeta$. However, for the PSPC gamma-ray model, double-peaked profiles occur only for large $\zeta$, because the $E_{||}$-dependence on $\phi_{\rm pc}$ and $\eta$ limits emission to favorably-curved field lines at high~$\alpha$. Therefore, an observer mostly sees emission from only one pole in this case, similar to the OG model.

Both OG and TPC models have a preponderance of double-peaked light curves at similar phases (see especially the lower right corners of Figures~\ref{fig:LC3} through~\ref{fig:LC7}). OG models do not exist at all angle combinations, while TPC models do (due to emission occuring below the null charge surface). It is interesting to note that one may find sharp, solitary peaks for some regions in phase space in OG models, while the corresponding TPC-peaks usually have additional low-level features (e.g., compare the TPC and OG profiles at $(\alpha,\zeta)=(30^\circ,60^\circ)$). Our models follow the inverse trend of peak separation vs.\ radio-to-gamma lag, first noticed by \citet{Romani95}. Our profile pulse width is proportional to $w$, because we assume that emission fills the full gap, unlike the case of \citet{Watters09}.

We chose best-fit light curves from the various models to match the MSP gamma-ray and radio data by eye, using plots such as those in Figures~\ref{fig:LC1} through~\ref{fig:LC7}. However, statistical uncertainties in the data may complicate unique matching of predicted and observed profiles. In addition, the model light curves usually do not radically change for a $\sim5^\circ$-change in $\alpha$ or $\zeta$, making our obtained fits somewhat subjective. The left panels of Figures~\ref{fig:LC_com1} through~\ref{fig:LC_com4} show phaseplots associated with the best light curve fits obtained for all eight MSPs (with horizontal lines indicating constant-$\zeta$ slices). In each case, the upper subpanel indicates a TPC model, and the lower one an OG model, except for PSR~J1744$-$1134 and PSR~J2124$-$3358, where the left panels are for PSPC models. We did not find any satisfactory fits from the geometric PC models, and TPC and OG models with $w=0$. In addition, the radio model fits the data quite well overall, except for the case of PSR~J0218+4232, which seem to require a wider cone beam.

In the right panels of Figures~\ref{fig:LC_com1} to~\ref{fig:LC_com4}, we show the observed gamma-ray and radio light curves, along with model fits. (We normalized the data to unity. Next, we normalized the model light curves to unity minus the background level. We lastly added this background to the latter.) Three MSPs (PSR~J0030+0451, PSR~J0218+4232, and PSR~J1614$-$2230) have double-peaked light curves, indicating the presence of screening electron-positron pairs (which are necessary to form the TPC or OG emitting structure). In six cases, the gamma-ray light curve lags the radio. Two MSPs, PSR~J0030+0451 and PSR~J1614$-$2230, have a relative phase lag $\delta\sim0.2$ (distinct from the function $\delta^\prime(\eta)$ used earlier), and four, PSR~J0218+4232, PSR~J0437$-$4715, PSR~J1613$-$0200, and PSR~J0751+1807, have $\delta\sim0.45$. These MSPs are well fit by TPC and OG models. The remaining two MSPs (PSR~J1744$-$1134 and PSR~J2124$-$3358) have $\delta\sim0.85$, which means that the radio lags the gamma-ray curves by 0.15 in phase. These two cases are exclusively fit by the PSPC model, where the gamma and radio emission come from the same magnetic pole, and originate well above the stellar surface. 
In the PSPC (and PC) model, the radio peak lags the gamma-ray peak, because the gamma-ray emission originates from all open field lines, appearing at earlier phases and washing out the caustic peaks.
For PSR~J1614$-$2230, the radio profile was measured at 1.5~GHz, and for PSR~J0437-4715, at 3~GHz (although for the modeling we only use frequencies 1.4~GHz and 3~GHz). All other radio profiles were observed at 1.4~GHz \citep{Abdo09_MSP}. Our best-fit model light curves allow us to infer values for $\alpha$ and $\zeta$ for each MSP. These are summarized in Table~\ref{tab2} (labeled with subscripts `TPC', `OG', and `PSPC'), and compared with values obtained from radio polarimetric measurements (labeled with subscripts `radio'). The latter inferred values are typically very difficult to obtain for MSPs due to the flatness of the position angle curve, and scatter of data. They are therefore generally quite uncertain.

We lastly calculated $f_\Omega(\alpha,\zeta)$ using Eq.~(\ref{eq:fom}) for each of the different models, and for different periods. Results are shown in Figures~\ref{fig:fom1} through~\ref{fig:fom4}. The `pinpoints' of more intense color which are sometimes visible is an artifact of our limited resolution of $5^\circ$ for $\alpha$ and $\zeta$. Note that very low-level emission at large impact angles may produce excessively large $f_\Omega$ factors, even for cases where the pulsar is not expected to be visible. For representational purposes, we set $f_\Omega=0$ when it exceeds the value of~4. We also calculated values for $f_\Omega$ for our best-fit models, and summarized them in Table~\ref{tab2}. Although $f_\Omega$ is a function of $\alpha$, $\zeta$, and $P$, it is typically of order unity for the best-fit geometries we consider here. The OG model typically predicts lower values than the TPC model. This is consistent with the findings of \citet{Watters09}. Although there are small differences when performing a detailed comparison of our results for TPC and OG models with those of \citet{Watters09}, our functional dependence of $f_\Omega(\alpha,\zeta)$ qualitatively resembles their results, and we obtain similar values of $f_\Omega(\alpha,\zeta)$ (keeping in mind that we are modeling MSPs, while they studied younger pulsars). Our results for the PSPC model however differ markedly from their PC model results, due to the fundamental physical difference of magnetospheric structure for MSPs and younger pulsars (i.e., unscreened vs.\ screened pulsar magnetospheres).

\section{Discussion and Conclusions}
\label{sec:Con}
We presented results from 3D emission modeling of gamma-ray and radio radiation in the framework of geometric PC, OG, and TPC pulsar models, and also for the full-radiation PSPC model. We have applied our results to recent measurements of newly-discovered MSPs by \textit{Fermi}-LAT. In this sense, we present results complementary to those obtained by \citet{Watters09} for young pulsars.

\begin{figure}
\epsscale{1.0}
\plotone{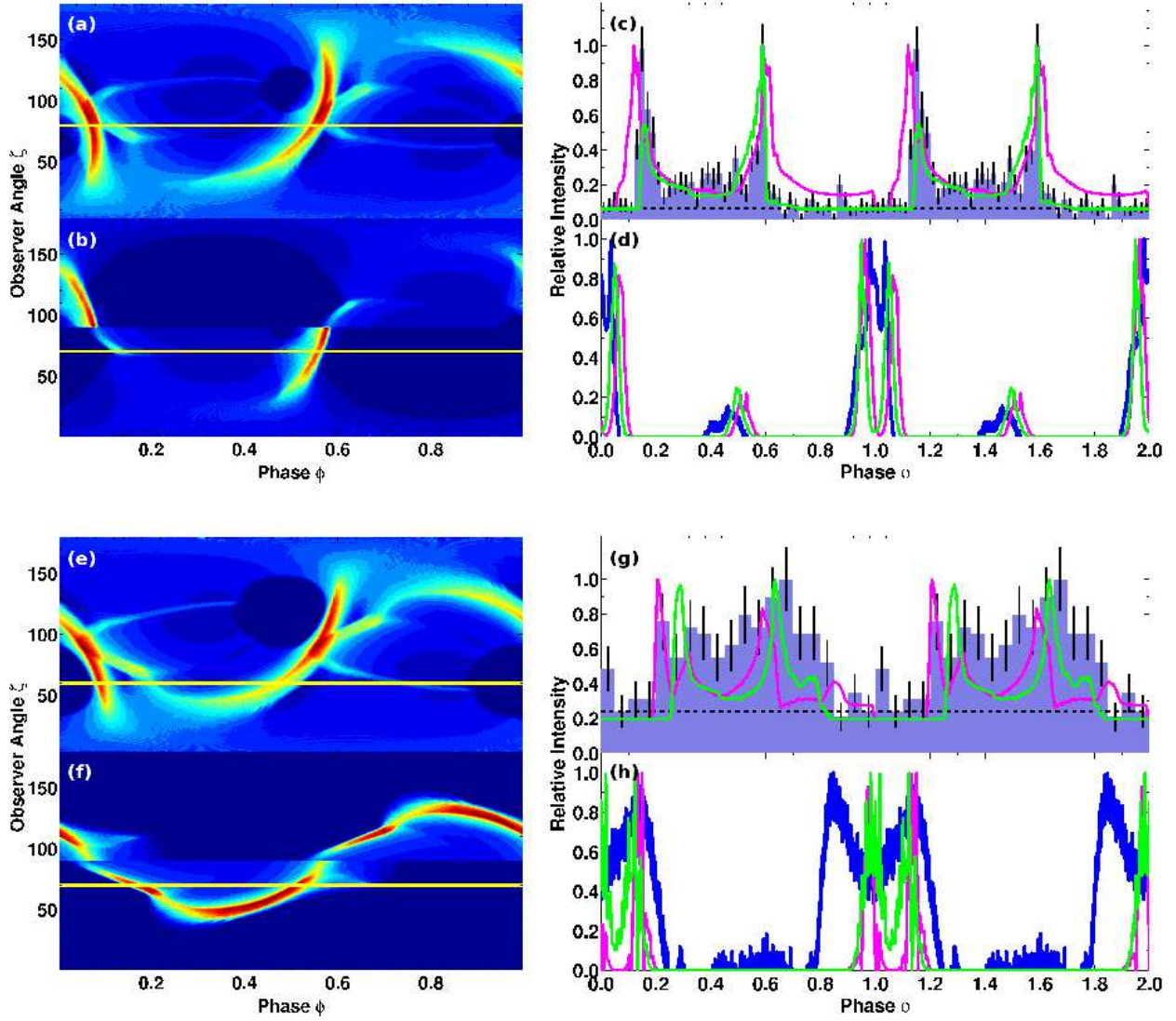}
\caption{Gamma-ray phaseplots (left panels) and observed and fitted gamma-ray and radio light curves (right panels) for PSR~J0030+0451 (panel [a]-[d], $P=5$~ms) and PSR~J0218+4232 (panel [e]-[h], $P=2$~ms). Panel~(a) is for a TPC1 model with $(\alpha,\zeta)=(70^\circ,80^\circ)$, (b) for an OG1 model with $(\alpha,\zeta)=(80^\circ,70^\circ)$, (e) for a TPC1 model with $(\alpha,\zeta)=(60^\circ,60^\circ)$, and (f) for an OG1 model with $(\alpha,\zeta)=(50^\circ,70^\circ)$.
For the gamma-ray light curves (panel [c] and [g]), the histograms represent the \textit{Fermi}-LAT data \citep{Abdo09_MSP}, the horizontal dashed line the estimated background level, the dashed (online: magenta) lines are TPC fits, and dash-dotted (online: green) lines are OG fits (see Table~\ref{tab2}). For the radio light curves (panel [d] and [h]) the solid (blue) line represents the radio data, while the dashed (magenta) and dash-dotted (green) lines correspond to the same $(\alpha,\zeta)$-combinations as those of the respective TPC and OG fits.\label{fig:LC_com1}}
\end{figure}

\begin{figure}
\epsscale{1.0}
\plotone{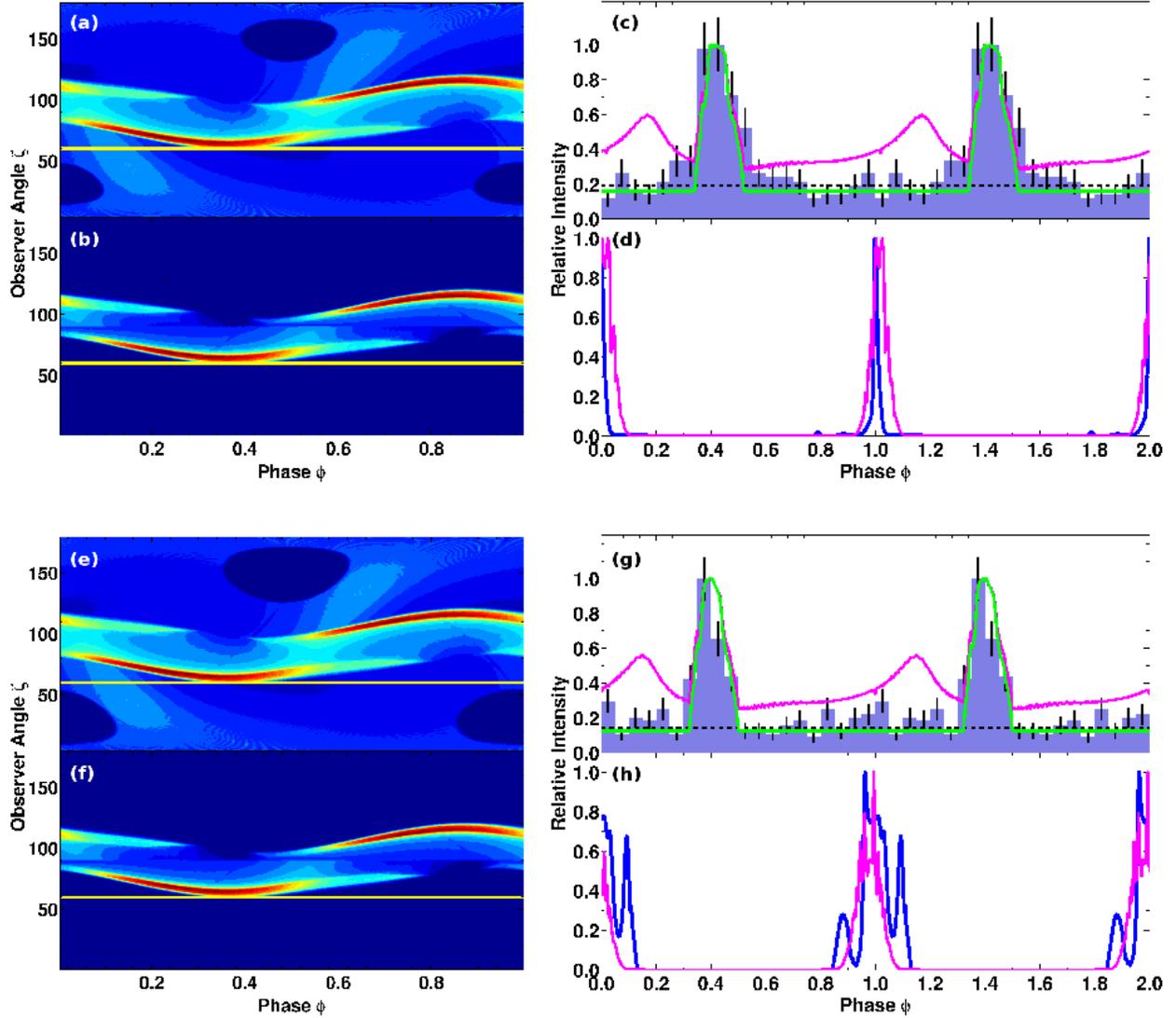}
\caption{Same as Figure~\ref{fig:LC_com1}, but for PSR~J0437-4715 (panel [a]-[d], $P=5$~ms) and PSR~J0613-0200 (panel [e]-[h], $P=3$~ms). Panel~(a) is for a TPC2 model with $(\alpha,\zeta)=(30^\circ,60^\circ)$, (b) for an OG1 model with $(\alpha,\zeta)=(30^\circ,60^\circ)$, (e) for a TPC2 model with $(\alpha,\zeta)=(30^\circ,60^\circ)$, and (f) for an OG1 model with $(\alpha,\zeta)=(30^\circ,60^\circ)$. For the cases where we use the same $(\alpha,\zeta)$-combination for both the TPC and OG fits, we only have a single radio light curve fit.\label{fig:LC_com2}}
\end{figure}

\begin{figure}
\epsscale{1.0}
\plotone{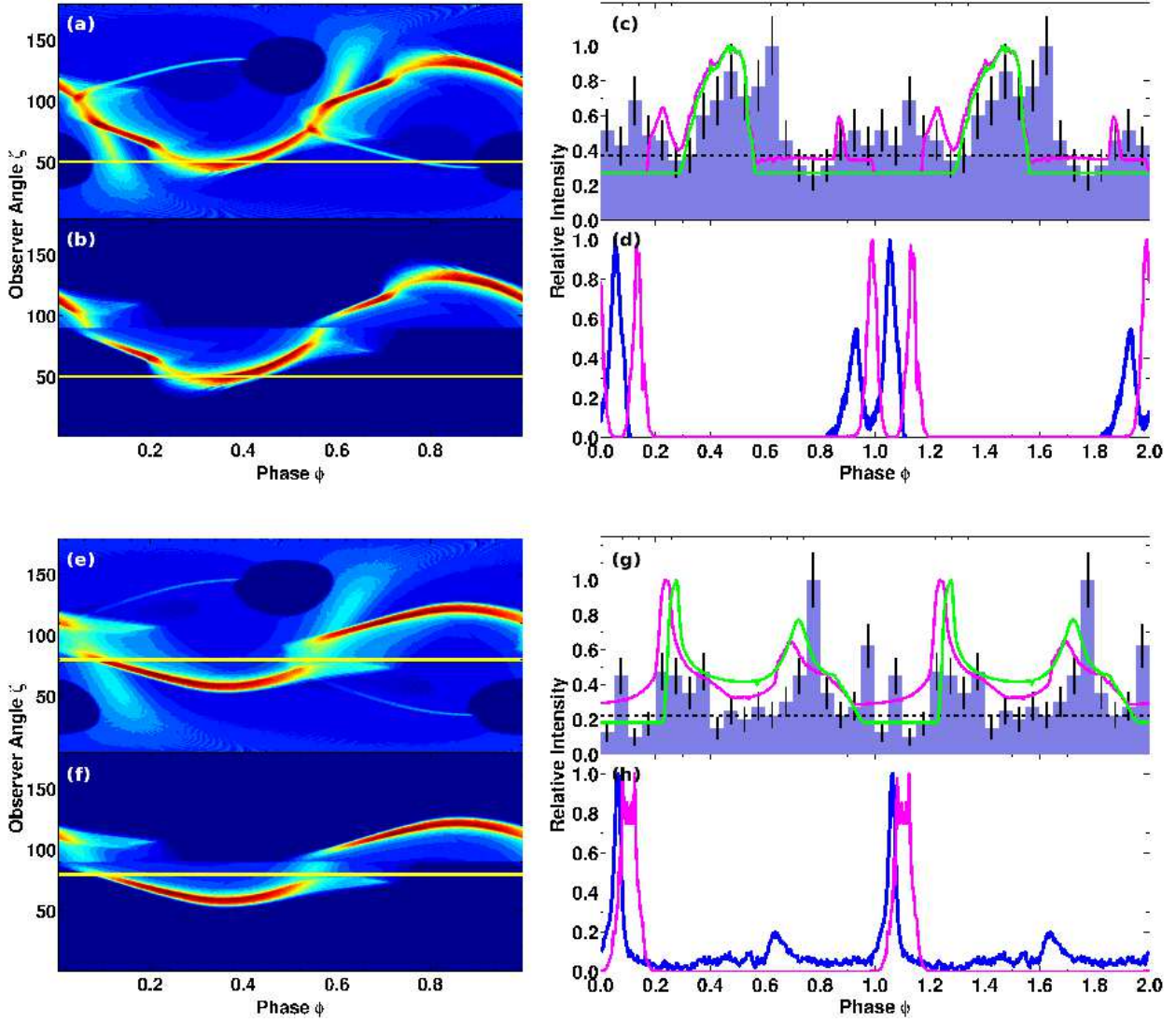}
\caption{Same as Figure~\ref{fig:LC_com1}, but for PSR~J0751+1807 (panel [a]-[d], $P=3$~ms) and PSR~J1614-2230 (panel [e]-[h], $P=3$~ms). Panel~(a) is for a TPC2 model with $(\alpha,\zeta)=(50^\circ,50^\circ)$, (b) for an OG1 model with $(\alpha,\zeta)=(50^\circ,50^\circ)$, (e) for a TPC2 model with $(\alpha,\zeta)=(40^\circ,80^\circ)$, and (f) for an OG1 model with $(\alpha,\zeta)=(40^\circ,80^\circ)$.\label{fig:LC_com3}}
\end{figure}

\begin{figure}
\epsscale{1.0}
\plotone{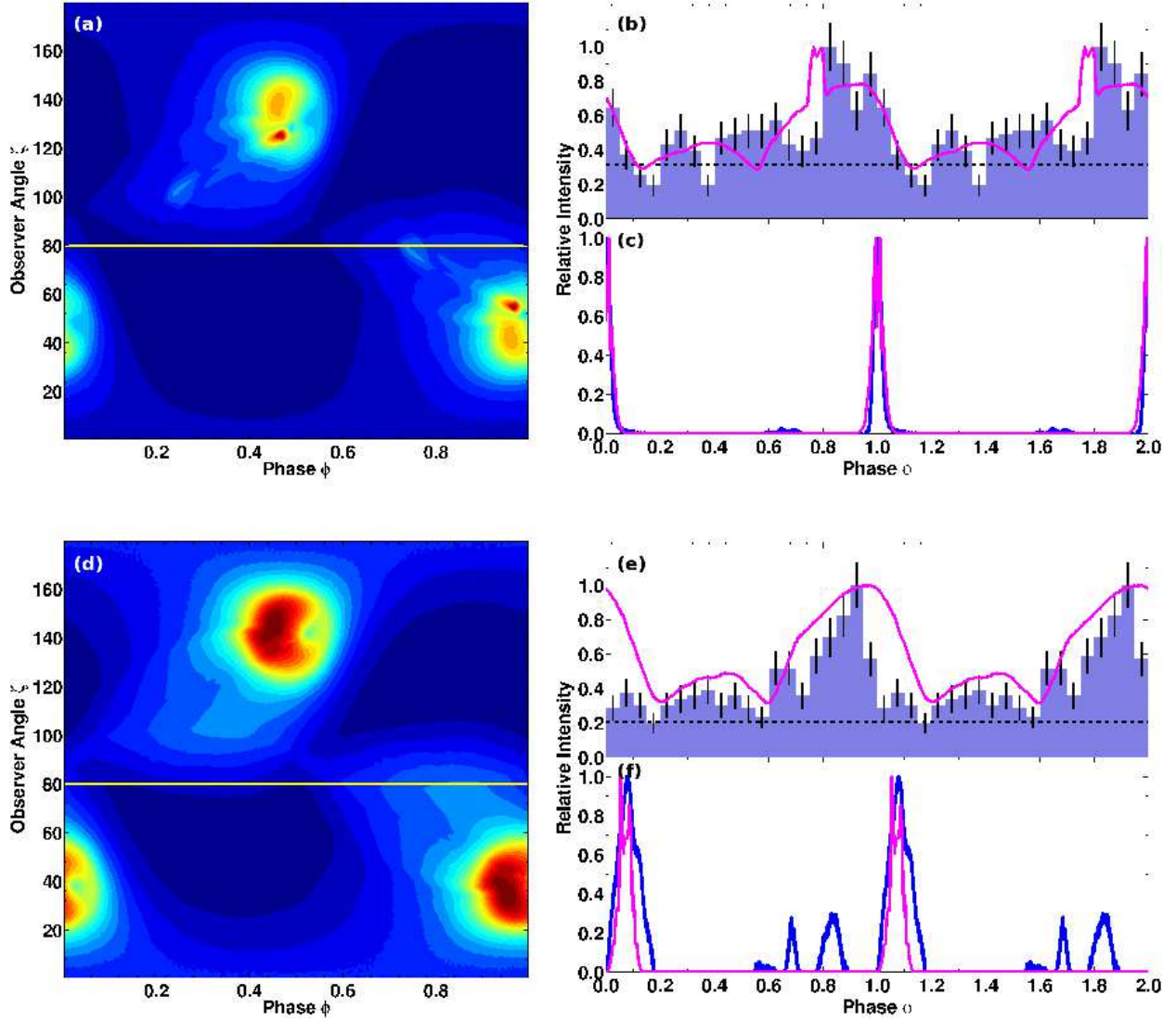}
\caption{Similar to Figure~\ref{fig:LC_com1}, but for PSR~J1744-1134 (panel [a]-[c], $P=5$~ms) and PSR~J2124-3358 (panel [d]-[f], $P=5$~ms). Panel~(a) is for PC2 model with $(\alpha,\zeta)=(50^\circ,80^\circ)$, and (d) for a PC2 model with $(\alpha,\zeta)=(40^\circ,80^\circ)$. In panels~(b) and~(e), the dashed (magenta) lines signify PC2 model fits.\label{fig:LC_com4}}
\end{figure}

\begin{figure}
\epsscale{0.8}
\plotone{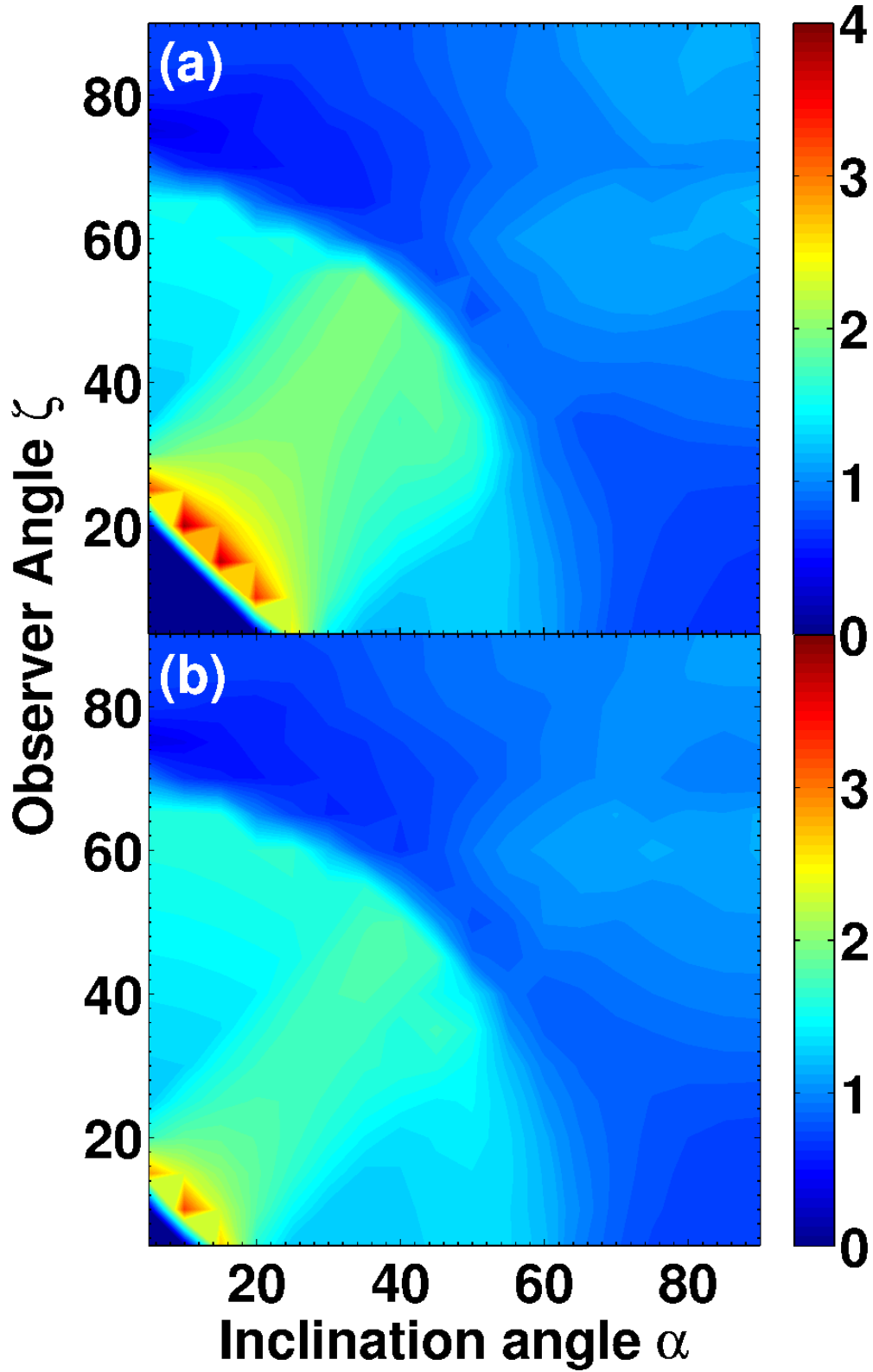}
\caption{The flux correction factor $f_\Omega(\alpha,\zeta,P)$ vs.\ $\alpha$ and $\zeta$ for a TPC2 model, with panel~(a) and~(b) for $P=2$~ms and $P=5$~ms, respectively. The same color scale is used throughout, and values of $f_\Omega>4$ were set to zero.\label{fig:fom1}}
\end{figure}

\begin{figure}
\epsscale{0.8}
\plotone{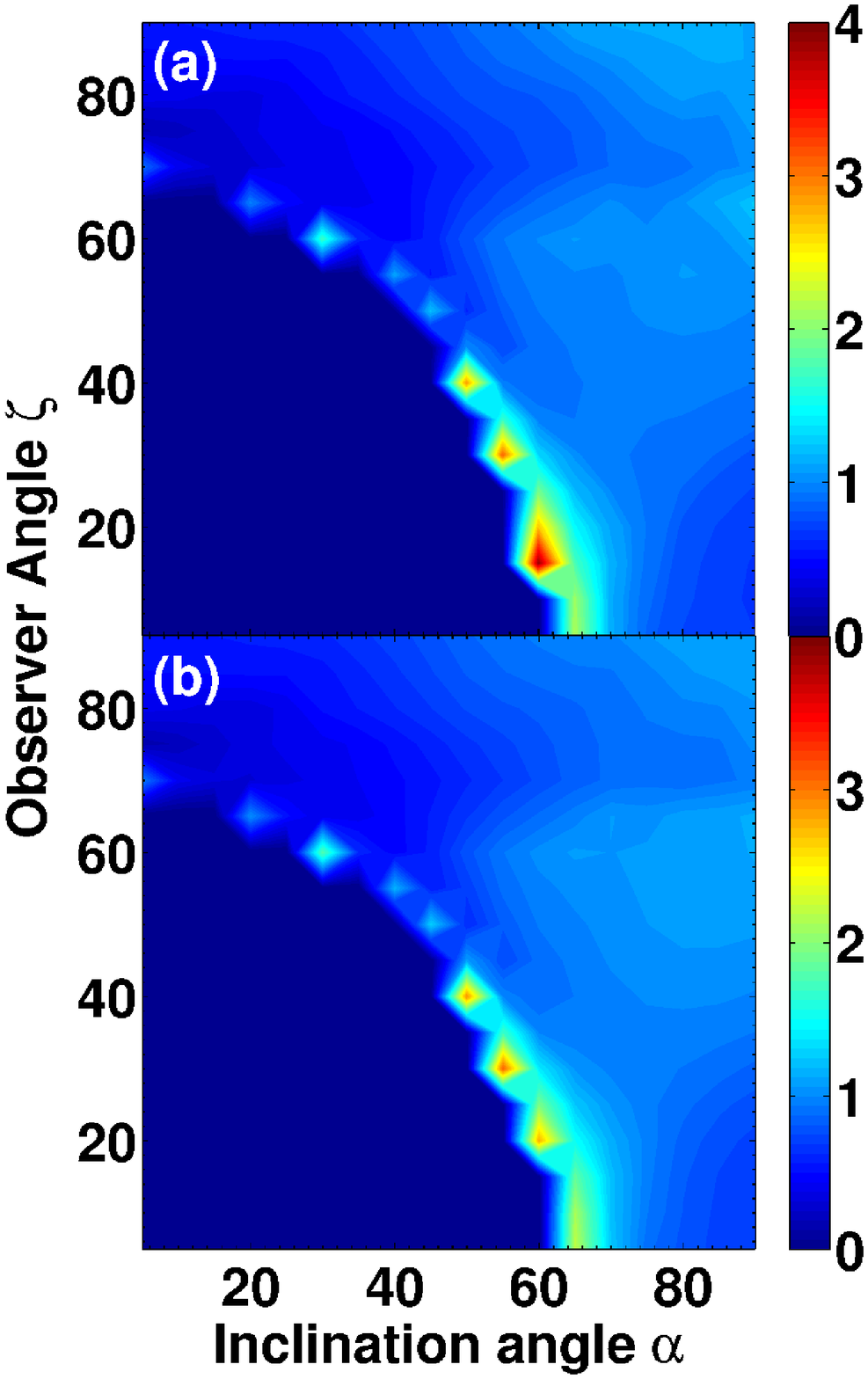}
\caption{The flux correction factor $f_\Omega(\alpha,\zeta,P)$ vs.\ $\alpha$ and $\zeta$ for an OG1 model, with panel~(a) and~(b) for $P=2$~ms and $P=5$~ms, respectively.\label{fig:fom2}}
\end{figure}

\begin{figure}
\epsscale{0.8}
\plotone{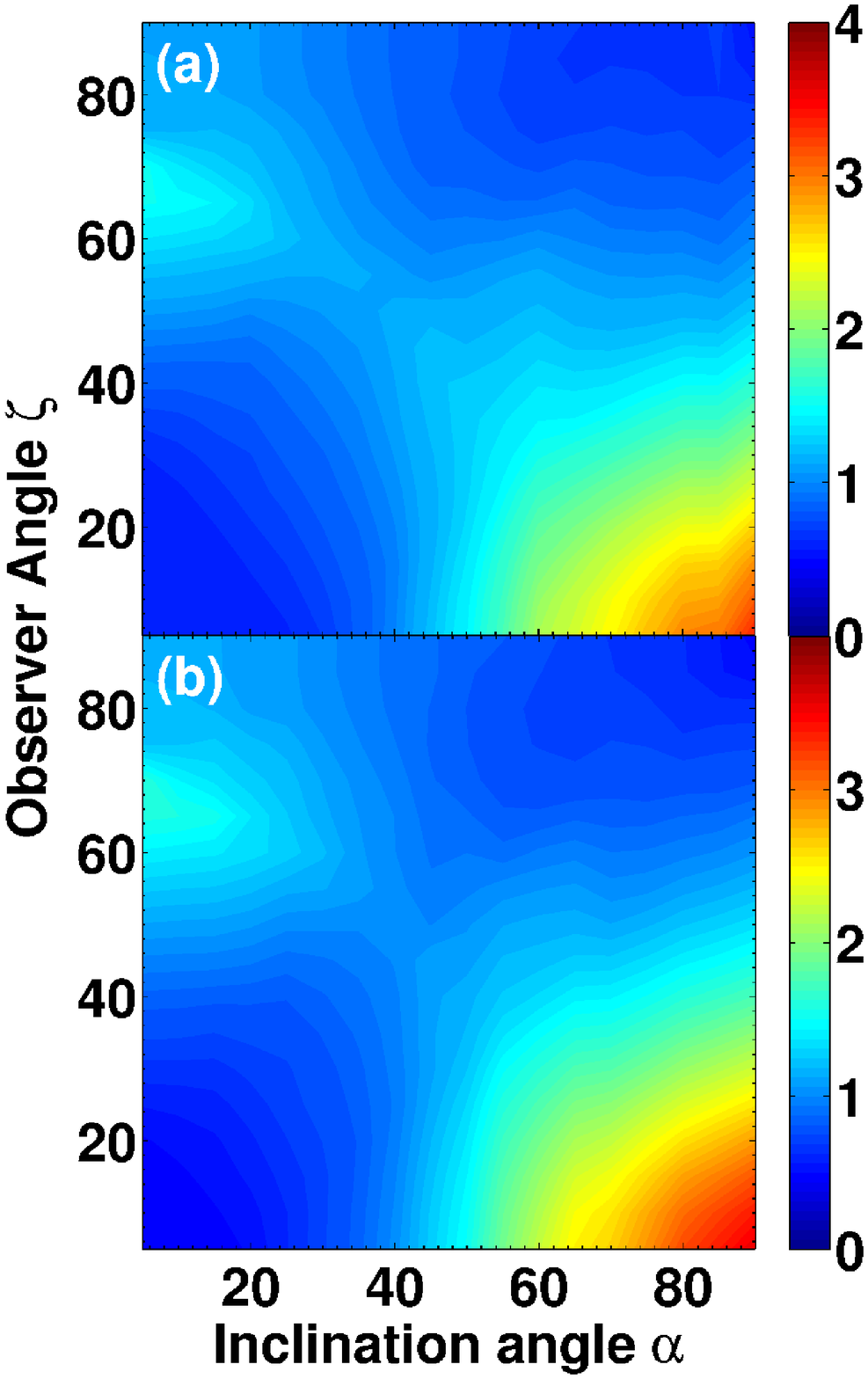}
\caption{The flux correction factor $f_\Omega(\alpha,\zeta,P)$ vs.\ $\alpha$ and $\zeta$ for a PC1 model, with panel~(a) and~(b) for $P=2$~ms and $P=5$~ms, respectively.\label{fig:fom3}}
\end{figure}

\begin{figure}
\epsscale{0.8}
\plotone{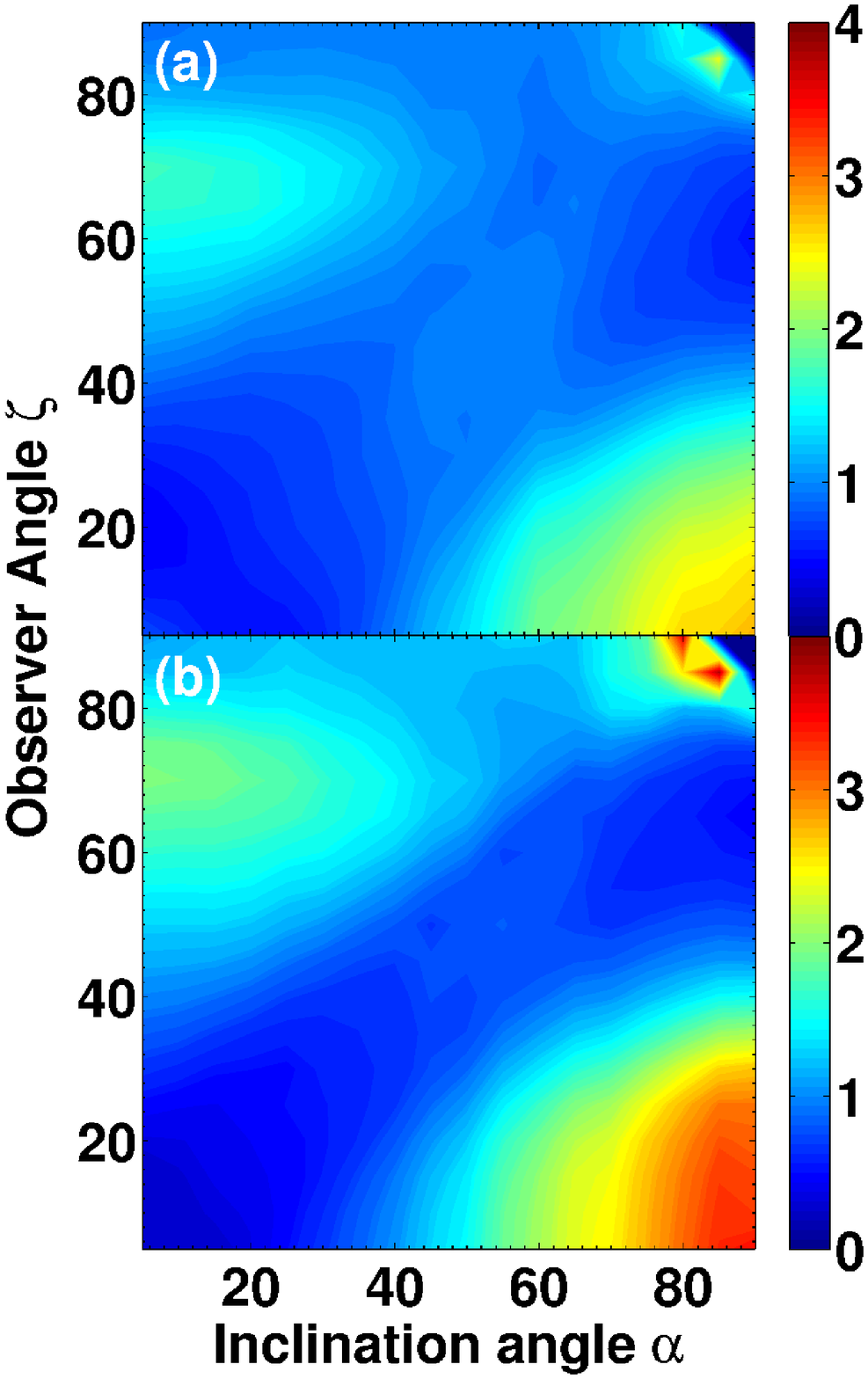}
\caption{The flux correction factor $f_\Omega(\alpha,\zeta,P)$ vs.\ $\alpha$ and $\zeta$ for a PC2 model, with panel~(a) and~(b) for $P=2$~ms and $P=5$~ms, respectively.\label{fig:fom4}}
\end{figure}

Previously, it was believed that most MSPs should have unscreened magnetospheres \citep{HUM05}, as they lie below the predicted CR pair death line on the $P\dot{P}$-diagram. It was expected that such pair-starved MSPs should have single gamma-ray pulses roughly in phase with the radio \citep{VdeJ05}. From Figure~\ref{fig:LC_com1} and~\ref{fig:LC_com3}, we see the surprising fact that there are indeed MSPs that have double-peaked light curves well fit by TPC / OG models, as are many of the young gamma-ray pulsars. This is interpreted as indicating the operation of a magnetic pair formation mechanism, and copious production of pairs to set up the required emitting gap structure. 

New ways of creating pairs in low-$\dot{E}_{\rm rot}$ pulsars will have to be found to explain this phenomenon. PSR~J0030+0451 illustrates this point very well in that it has the lowest $\dot{E}_{\rm rot}$ of the MSP sample ($3.5\times10^{33}$~erg s$^{-1}$), therefore lying significantly below the calculated CR death line \citep[e.g.,][]{HMZ02}, and yet exhibits the sharpest double peaks of the current population, implying emission originating in very thin TPC / OG gaps. The problem may be alleviated somewhat by increasing the stellar compactness $\kappa^\prime$ (larger mass or smaller radius), motivated by recent measurements of large MSP masses \citep[up to $\sim1.7M_\odot$; see][and references therein]{Verbiest08,Freire09}. This will boost the GR E-fields, and enhance pair creation probability. Another way to do this would be to increase the magnetic field. B-fields that are larger than those usually inferred using the dipole spin-down model (and having smaller curvature radii) may be present when there are multipolar B-components near the surface (or an offset-dipole geometry).  In fact, offset dipoles have been suggested in modeling the X-ray light curves of MSPs J0437$-$4715 and J0030$+$0451 \citep{Bogdanov07,Bogdanov09}.  However, detailed investigation of such a scenario and its implications for pair cascades is necessary to place this speculation on sure footing. Another possible origin for higher surface fields is the movement of magnetic poles toward the spin axis during the spin-up phase of an MSP \citep{Lamb08}. It has been argued that during the spin-up to millisecond periods, the inward motion of the neutron star superfluid vortices produces a strain on the crust, causing the magnetic poles to drift toward the spin-axis \citep{Ruderman91}. If the two poles are in the same hemisphere prior to spin-up, the poles drift toward each other, producing a nearly orthogonal rotator having the same dipole moment but a surface field that can be orders of magnitude higher \citep{Chen93}.

We find that there is exclusive differentiation between the TPC / OG models on the one hand, and the PSPC model on the other hand. Six MSPs have gamma-ray light curves which lag the radio and are explained using TPC or OG fits, but not PSPC fits. For the remaining two MSPs, the radio light curves slightly lag the gamma-ray light curves, and these are fit by the PSPC model (and not by the TPC / OG models). It therefore seems that there are two subclasses emerging within the current gamma-ray MSP sample, and it is not obvious which pulsar characteristics provide a means to predict subclass membership. From our model light curve fitting, we furthermore find $(\alpha,\zeta)$ values which are in reasonable agreement with values inferred from MSP polarization measurements. Although we find good PSPC fits for the last two MSPs, we caution that the E-field is only approximately known (e.g., it follows from a local electrodynamical model based on a GR dipolar B-field). Future models which take global current flow patterns into account, along with more sophisticated B-field structure, may produce more realistic solutions for the E-field.

Our ability to discriminate between different classes of models derives from the fact that we produced both the gamma-ray and radio curves within the same model. We could then use the shape \textit{and} relative radio-to-gamma phase lag provided by the data to obtain the best-fit model type for each MSP. The data also enabled us to conclude that the emission, in \textit{all} models considered, must come from the outer magnetosphere. This has now been observed to be true for the bulk of the gamma-ray pulsar population \citep{Abdo09_Cat}.

In the case of PSR~J0437-4715 and PSR~J0613-0200, we find that the TPC model predicts a significant precursor to the main gamma-ray peak, while the OG model predicts no such low-level emission. With more statistics, this effect may possibly become a discriminator between the TPC and OG models. (We assumed that the TPC emission region starts at $r_{\rm em}=R$ when creating our plots. However, the relative intensity of the precursor and low-level emission predicted by the TPC model may be reduced by limiting the emission region's extension, i.e.\ only collecting photons above a certain radius $r_{\rm em} \geq R_{\rm min} > R$.)

We calculated the flux correction factor in the context of the different models, and found that $f_\Omega \sim1$. These values imply a wide beaming angle, and derives from the fact that we obtain best fits for large impact angles. \citet{Venter_phd} previously found $\Lambda_{\rm avg} \sim 10-30$ (i.e.\ $f_\Omega\sim0.8-2.4$), and $\Lambda_{\max} \sim300$ ($f^{\max}_\Omega\sim24$) for the PSPF model. Now, we find $f_\Omega$ $\sim 0.5-2$, and $f^{\max}_\Omega\sim 4$. These results are roughly consistent, with the differences stemming from the following: (i) \citet{Venter_phd} used energy flux ratios to calculate $\Lambda$, while we are using photon flux ratios to calculate $f_\Omega$, assuming that the photon and energy fluxes have similar distributions across $(\zeta,\phi)$-space; (ii) \citet{Venter_phd} only used $E_{||}^{(1)}$ and $E_{||}^{(2)}$ for the E-field, while we now also include the high-altitude solution ($E_{||}^{(3)}$) for the PSPF case. This leads to more intense high-altitude emission, and therefore smaller values of $f_\Omega$ for off-beam emission.

We lastly remark that the larger radio beam widths of MSPs compared to those of canonical pulsars should lead one to expect relatively few radio-quiet MSPs.

The spectacular data from \textit{Fermi}-LAT hold the promise of phase-resolved spectroscopy, at least for the brightest pulsars, and will challenge existing pulsar models to reproduce such unprecedented detail. Future work therefore includes using full acceleration and radiation models to study gamma-ray spectra, luminosities, and light curves, in order to constrain fundamental electrodynamical quantities, and possibly providing the opportunity of probing the emission geometry and B-field structure more deeply. Improved understanding of pulsar models will also feed back into more accurate population synthesis models \citep[e.g.,][]{Story07}. In addition, we hope to obtain better understanding of important quantities such as MSP efficiencies, and whether this quantity is similar for Galactic-Field and globular-cluster MSPs \citep{Abdo09_Tuc}.

\acknowledgments
CV is supported by the NASA Postdoctoral Program at the Goddard Space Flight Center, administered by Oak Ridge Associated Universities through a contract with NASA, and also by the South African National Research Foundation. AKH acknowledges support from the NASA Astrophysics Theory Program. We thank Alex Muslimov and Jarek Dyks for useful discussions.

\clearpage
\begin{deluxetable}{ccccccc} 
\tablecolumns{7} 
\tablewidth{0pc} 
\tablecaption{Parameters of MSPs discovered by \textit{Fermi}-LAT \citep{Abdo09_MSP}\label{tab1}} 
\tablehead{ 
\colhead{Name} & \colhead{$P$} & \colhead{$\dot{P}$} & \colhead{Distance} & \colhead{Age} & \colhead{$\dot{E}_{\rm rot}$} & \colhead{$B_0$}\\
\colhead{} & \colhead{(ms)} & \colhead{($10^{-20}$)} & \colhead{(kpc)} & \colhead{($10^9$ yr)} & \colhead{($10^{33}$~erg s$^{-1}$)} & \colhead{($10^8$~G)}
}
\startdata 
J0030+0451   & 4.865 & 1.01  & 0.300 $\pm$ 0.090  & 7.63 & 3.47 & 2.04\\
J0218+4232   & 2.323 & 7.79  & 2.70  $\pm$ 0.60   & 0.47 & 245 & 4.31\\
J0437$-$4715 & 5.757 & 1.39  & 0.156 $\pm$ 0.002  & 6.55 & 2.88 & 2.87\\
J0613$-$0200 & 3.061 & 0.915 & 0.48  $\pm$ 0.14   & 5.31 & 12.6 & 1.69\\
J0751+1807   & 3.479 & 0.755 & 0.62  $\pm$ 0.31   & 7.30 & 7.08 & 1.64\\
J1614$-$2230 & 3.151 & 0.397 & 1.30  $\pm$ 0.25   & 12.6 & 5.01 & 1.13\\
J1744$-$1134 & 4.075 & 0.682 & 0.470 $\pm$ 0.090  & 9.47 & 3.98 & 1.69\\
J2124$-$3358 & 4.931 & 1.21  & 0.25  $\pm$ 0.13   & 6.47 & 3.98 & 2.47\\
\enddata 
\end{deluxetable}

\clearpage
\begin{deluxetable}{cccccc} 
\tablecolumns{5} 
\tablewidth{0pc} 
\tablecaption{MSP Model Descriptions\label{tab3}}
\tablehead{ 
\colhead{Abbreviation} & \colhead{$r_{\rm ovc}$} & \colhead{$w$} & \colhead{$\delta r_{\rm ovc}$} & \colhead{Azimuthal bins} & \colhead{Description}
}
\startdata 
TPC1 & $[0.90,1.00]$ & 0.10 & 0.005 & 180 & Geometric TPC Model\\
TPC2 & $[0.95,1.00]$ & 0.05 & 0.005 & 180 & Geometric TPC Model\\
TPC3 & $[0.80,1.00]$ & 0.20 & 0.005 & 180 & Geometric TPC Model\\
TPC4 & $[0.60,1.00]$ & 0.40 & 0.005 & 180 & Geometric TPC Model\\
TPC5 & $[1.00,1.00]$ & 0.00 & 0.005 & 180 & Geometric TPC Model\\
OG1  & $[0.95,1.00]$ & 0.05 & 0.005 & 180 & Geometric OG Model\\
OG2  & $[0.90,0.90]$ & 0.00 & 0.005 & 180 & Geometric OG Model\\
OG3  & $[1.00,1.00]$ & 0.00 & 0.005 & 180 & Geometric OG Model\\
PC1  & $[0.00,1.00]$ & 1.00 & 0.005 & 180 & Geometric PC Model\\
PC2  & $[0.00,1.00]$ & 1.00 & 0.005 & 180 & Radiation PSPC Model\\
\enddata 
\end{deluxetable} 

\clearpage
\begin{deluxetable}{ccccccccccccc} 
\tabletypesize{\scriptsize}
\tablecolumns{13} 
\tablewidth{0pc} 
\tablecaption{Model fits for $\alpha$, $\zeta$, and $f_\Omega(\alpha,\zeta,P)$\label{tab2}} 
\tablehead{ 
\colhead{Name} & \colhead{$\alpha_{\rm TPC}$} & \colhead{$\zeta_{\rm TPC}$} & \colhead{$\alpha_{\rm OG}$} & \colhead{$\zeta_{\rm OG}$} & \colhead{$\alpha_{\rm PSPC}$} & \colhead{$\zeta_{\rm PSPC}$} & \colhead{$\alpha_{\rm radio}$} & \colhead{$\zeta_{\rm radio}$} & \colhead{Ref.} & \colhead{$f_{\rm \Omega,TPC}$} & \colhead{$f_{\rm \Omega,OG}$} & \colhead{$f_{\rm \Omega,PSPC}$}\\
\colhead{} & \colhead{($^\circ$)} & \colhead{($^\circ$)} & \colhead{($^\circ$)} & \colhead{($^\circ$)} & \colhead{($^\circ$)} & \colhead{($^\circ$)} & \colhead{($^\circ$)} & \colhead{($^\circ$)} & \colhead{} & \colhead{} & \colhead{} & \colhead{}
}
\startdata 
J0030+0451   & 70      & 80      & 80      & 70      & \nodata & \nodata & $\sim62$      & $\sim72$          & 1   & 1.04    & 0.90    & \nodata\\
J0218+4232   & 60      & 60      & 50      & 70      & \nodata & \nodata & $\sim8$       & $\sim90$          & 2   & 1.06    & 0.63    & \nodata\\
J0437$-$4715 & 30      & 60      & 30      & 60      & \nodata & \nodata & $20-35$       & $16-20$           & 3,4 & 1.23    & 1.82    & \nodata\\
J0613$-$0200 & 30      & 60      & 30      & 60      & \nodata & \nodata & small $\beta$ & \nodata           & 5   & 1.19    & 1.76    & \nodata\\
J0751+1807   & 50      & 50      & 50      & 50      & \nodata & \nodata & \nodata       & \nodata           &     & 0.80    & 0.65    & \nodata\\
J1614$-$2230 & 40      & 80      & 40      & 80      & \nodata & \nodata & \nodata       & \nodata           &     & 0.83    & 0.64    & \nodata\\
J1744$-$1134 & \nodata & \nodata & \nodata & \nodata & 50      & 80      & \nodata       & \nodata           &     & \nodata & \nodata & 1.19\\
J2124$-$3358 & \nodata & \nodata & \nodata & \nodata & 40      & 80      & $20-60$ (48)  & $27-80$ (67)      & 6   & \nodata & \nodata & 1.29\\
\enddata 
\tablerefs{
(1) \citet{Lommen00};
(2) \citet{Stairs99};
(3) \citet{MJ95};
(4) \citet{Gil97}, 
(5) \citet{Xilouris98};
(6) \citet{Manchester04}}
\end{deluxetable}

\clearpage

\clearpage

\clearpage

\clearpage

\clearpage

\clearpage

\clearpage

\clearpage

\clearpage

\clearpage

\clearpage

\clearpage

\clearpage

\clearpage

\clearpage

\clearpage

\clearpage

\clearpage

\clearpage

\clearpage

\clearpage

\clearpage

\clearpage

\end{document}